\documentclass{aa}  

\usepackage{graphicx}
\usepackage{txfonts}
\usepackage{multirow}
\usepackage{makecell}

\usepackage{amsmath}	% Advanced maths commands
\usepackage{amssymb}	% Extra maths symbols
\usepackage{physics}
\usepackage{bm} 
\usepackage[normalem]{ulem}  % or striking through

\newcommand{\glcomment}[1]{} 
\newcommand{\comment}[1]{} % for not displaying some parts of the text

\newcommand{\RISCO}{R_{\rm ISCO}}
\newcommand{\rc}{r_{\rm c}}
\newcommand{\rl}{r_{\rm l}}
\newcommand{\omegas}{\Omega_{\rm s}}
\newcommand{\omegak}{\Omega_\mathrm{K}}

\newcommand{\ra}{r_{\rm a}}
\newcommand{\fim}{\varphi} % just short for \varphi
\newcommand{\pd}[2]{\frac{\partial #1}{\partial #2}} %partial derivation

\newcommand{\Mdot}{\dot M}
\newcommand{\Mdotpr}{\dot{M}_\mathrm{prop}}

\newcommand{\akglOK}[1]{\textcolor{black}{#1}}
\newcommand{\dif}{\mathrm{d}}
\newcommand{\Tc}{T_{\rm c}}
\newcommand{\Teff}{T_{\rm eff}}
\newcommand{\Sc}{\Sigma}
\newcommand{\Wrf}{W_{r\varphi}}
\newcommand{\alphadisc}{\texttt{AlphaDiSC}~}

\usepackage{xcolor}

\renewcommand{\vb}[1]{\bm{#1}}

\usepackage{newtxtext,newtxmath}
\begin{document}

%%%%%%%%%%%%%%%%%%%%%%%%%%%%%%%%%%%%%%%%
% if you use custom commands in your title,
% ensure to check your title when submitting!
%%%%%%%%%%%%%%%%%%%%%%%%%%%%%%%%%%%%%%%%
   \title{Radial and vertical structure of accretion discs around magnetized stars} 

   %\subtitle{Subtitle}

%%%%%%%%%%%%%%%%%%%%%%%%%%%%%%%%%%%%%%%%
% Please separate each author with the \and command
%
% Use the \corrauth to provide the corresponding
% author address. It will be automatically inserted as 
% footnote in the PDF output.
%
% Please DO NOT include ORCIDs next to author names.
% Instead, please provide an active address for each coauthor:
% it will be automatically extracted by EDPS editorial system, 
% and co-authors will be be able to authenticate their ORCID.
%
% Only authenticated ORCIDs will be taken into account.
% ORCIDs included here will be removed.
%%%%%%%%%%%%%%%%%%%%%%%%%%%%%%%%%%%%%%%%

   \author{A. Kuzin\inst{1}\corrauth{\\ aleksei.kuzin@astro.uni-tuebingen.de}      
        \and G. Lipunova\inst{2, 3}\email{gvlipunova@gmail.com}
        }

   \institute{Institut f\"ur Astronomie und Astrophysik, Universit\"at T\"ubingen, Sand 1, D-72076 T\"ubingen, Germany
   \and Dr. Karl Remeis-Observatory and Erlangen Centre for Astroparticle Physics, Friedrich-Alexander Universität Erlangen-Nürnberg,
    Sternwartstr. 7, 96049 Bamberg, Germany
    \and Max Planck Institut f\"ur Radioastronomie, Auf dem H\"ugel 69, D-53121 Bonn, Germany }

   \date{Received September 30, 20XX}

% \abstract{}{}{}{}{}
% 5 {} token are mandatory
   \abstract
  % context heading (optional)
  % {} leave it empty if necessary  
   {The ubiquity of accretion discs around magnetized stars 
motivates studies of the structure of these discs, in particular in whether the external magnetic field changes the radial and vertical profiles of pressure, temperature, and other quantities. }
  % aims heading (mandatory)
   {We study the effects of the magnetic field of a central star on a steady accretion disc under the assumption that the magnetic field of the central star is able to partially penetrate the disc, with a degree of penetration quantified by a single parameter.} 
  % methods heading (mandatory)
   {We construct an analytical model of the interaction between the disc and the star's magnetic field.
The equations of motion are modified to take the magnetic field into account.
We solve the vertical structure of the disc assuming tabulated opacities and incorporate the magnetic field in order to determine the radial and vertical distributions of the disc parameters.
}
  % results heading (mandatory)
   {The inner radius of the accretion disc, the viscous stress tensor, and the spin-up torque on the central object follow from the obtained distribution of the magnetic torque. 
   In the limit where the stellar field is expelled from the disc, the inner radius scales with the Alfven radius, whereas when the external field penetrates the disc, the inner radius behaves in a more complex way, remaining within a few corotation radii.
The normalized vertical profile depends only weakly on the radius and magnetic field.
Nevertheless, the magnetic field can significantly modify the viscous stress tensor, radial structure, spectrum of the disc, and its thermal-equilibrium curves.
The publicly available code \alphadisc now includes discs around magnetized stars.
%The model is implemented in the publicly available code \alphadisc.
}
  % conclusions heading (optional), leave it empty if necessary
   {}


   \maketitle
    \nolinenumbers

%%%%%%%%%%%%%%%%%%%%%%%%%%%%%%%%%%%%%%%%%%%%%%%%%%%%%%%%%%%%%%
\section{Introduction}

Accretion discs surrounding magnetized stars are found in various astrophysical sources. Changes in the accretion rate are believed to be the underlying cause of phenomena such as bursts in dwarf novae \citep[e.g.][]{Livio-Pringle1992, Hameury20}, X-ray transients \citep[e.g.][]{KingRitter98, Lip22, Coban24}, and young stellar objects \citep[YSOs, see e.g.][]{BellLin94}. Furthermore, the interaction between the discs and the stars' magnetospheres plays a critical role in modulating the stars' spin periods \citep[e.g.][]{DangeloSpruit2011, ABP24, Sti25}. 
Despite the significant amount of analytical work on the interaction between accretion discs around magnetic stars and the stars' magnetic fields \citep[to name a few,][]{GL79a, GL79b, Aly80, AlyKuj90, Campbell92, Lovelace95, Wang95, Wang97, MP05, KR07, Begel15, Chash17, Chash19}, it is often assumed that the influence of the stellar magnetic field is limited to setting the inner disc radius, while the remaining structure and emission properties of the disc remain unchanged from the standard non-magnetic case.

In this paper, we investigate the radial and thermal structure of a stationary accretion disc interacting with a magnetized star.
The interaction between the disc and the star's magnetosphere is a complex problem that has been the focus of extensive research.
Two theoretical approaches exist in this regard.
The first suggests that the accretion disc behaves as a superconductor; this is justified  by the extremely high conductivity for the ionized medium.
Consequently, the field is completely expelled from the disc, and the resulting configuration can be approximated by the solution of \citet{Aly80}. 
The second approach assumes that the external magnetic field penetrates the disc~\citep[e.g. via Kelvin-Helmholtz instabilities and turbulent diffusion,][]{GL79a}, allowing the dipole field to thread the disc, possibly with a screening factor~\citep{AlyKuj90, Campbell92, Wang95, Wang97, KR07}. Some works suggest that the magnetic field penetrates the disc only in a narrow zone near the inner radius \citep{GL77, GL79a, GL79b, MP05},
while others \citep[e.g.][]{Lai99} suggest wide interaction zone, but that the degree of magnetic field penetration into the disc is determined by the temporal variability of the field.

In this work, we aim to reconcile these somewhat opposing approaches. We assume that the magnetic field can partially penetrate the disc. To quantify this effect, we introduce a parameter, $\eta$, defined as the fraction of the stationary vertical component of the dipole magnetic field that penetrates the disc.

If the external magnetic field partially threads the disc, we assume that the magnetic field lines are frozen into the plasma and entrained by the rotating matter, thereby inducing toroidal field components.

Determining the position of the inner disc radius is a non-trivial task and usually involves additional physical assumptions. The inner radius of an ideally conducting disc is often associated with the Alfven radius, which is the characteristic radius of a magnetosphere surrounded by radially infalling matter~\citep{Lamb+1973}.
When the magnetic field penetrates the disc, it acts as an agent for transferring the angular momentum of the matter in the radial direction, and the position of the inner radius must therefore be determined consistently with this magnetic torque.

For example, \citet{KR07} considered a `transition zone' between the star and the disc, and a solution of the angular momentum transport equation in this zone allowed the authors to determine the inner disc radius. We do not consider such a transition zone; instead, based on their algorithm, we develop a general approach for determining the position of the inner radius of the disc in the case of a partially penetrating magnetic field, parametrized by the parameter $\eta$. 

The radial distribution of the viscously released heat also depends on the parameter $\eta$.
The interaction of the disc with the magnetic field generally leads to changes in the thermal structure of the disc and subsequently influences the thermal stability of the disc's vertical structure. We investigate, for the first time, how the interaction with the magnetic field changes the critical conditions for the instability associated with the partial ionization of hydrogen in a disc around a magnetized star.

The numerical results presented in this work were obtained using the code \texttt{AlphaDiSC} \citep[][hereafter TLM23]{Tavl23}, which we modified to include the effects of the magnetic field. The modified code is open-source and available on GitHub\footnote{\url{https://github.com/AndreyTavleev/AlphaDiSC}}. The original \texttt{AlphaDiSC} was designed to calculate the vertical structure of $\alpha$-discs not embedded in an external magnetic field. The code also calculates the radial structure and S-curves. It utilizes realistic numerical opacities and equation of state, which allowed us to study the influence of the external magnetic field on the thermal-equilibrium states of  partially ionized discs in a more straightforward way than in previous numerical studies \citep{Scepi18, Scepi19}. In the following sections, we discuss in detail the differences in vertical structure calculations between the non-magnetized and magnetized cases.

In our numerical results, we assume parameters relevant to neutron stars (NSs) and white dwarfs. Discs around magnetized NSs are found in Accreting Millisecond X-ray Pulsars (AMXPs) and X-ray pulsars (XRPs). The relatively weak magnetic fields of AMXPs, $B \sim 10^{8}$~G, allow the disc to extend almost to the stellar surface, resulting in very bright disc emission during type II \hbox{X-ray} outbursts. This makes AMXPs suitable objects for studying discs interacting with magnetic fields.
Additionally, the magnetic field of the central star may significantly influence the structure of the accretion discs in intermediate polars (IPs) and T Tauri stars.

The paper is organised as follows. In Section \ref{sec:model}, we describe the accretion disc model. In Section \ref{sec:mag_field}, we discuss previous models of the disk-magnetosphere interaction and present our own model. Section \ref{sec:method} is devoted to the method used to calculate the radial and vertical structure. In Section \ref{sec:results}, we present the results for the vertical and radial structure, as well as the S-curves. We discuss our results in Section \ref{sec:colcl} and summarise the findings of the paper in Section \ref{sec:summary}.

\section{Model of the disc}\label{sec:model}

Throughout this work, we use cylindrical coordinates ($r, \fim, z$) in a non-rotating reference frame. The star is located at $r = z = 0$, and the equatorial plane of the disc lies at $z = 0$.
The rotation axis of the central star is aligned with the axis of the accretion disc.
We assume that the magnetic field of the star is dipolar.
The inclination angle $\chi$ between the magnetic moment $\bm{\mu}$ and the angular velocity of the star $\vb{\Omega}_s$ lies between $0^{\circ}$ and $90^{\circ}$.
The model is not applicable beyond the light cylinder radius: there is no dipolar field there, nor are there any field lines connecting the star and the disc (see Sec.~\ref{sec:mag_field} for details).

Within the light cylinder, matter within (outside) the geometrically thin disc is assumed to rotate with the Keplerian (stellar) angular velocity.
Generally speaking, the angular velocity of the matter in the disc differs from the Keplerian value because of the presence of the magnetic field; see \S~\ref{s.main_eqs} below.

\subsection{Notation}\label{sec:notation}
The following variables will be used throughout the paper.
The corotation radius is the distance at which the Keplerian angular velocity equals that of the star $\omegas$:
\begin{equation}\label{eq:corot}
 \rc  = \left(\frac{GM}{\omegas^2}\right) ^ {1/3}.
\end{equation}

The inner radius of the disc is denoted by $r_0$. The fastness parameter $\omega$ is defined as the ratio of the star's angular velocity $\omegas$ to the Keplerian angular velocity at $r_0$:
\begin{equation}
  \omega = \frac{\omegas}{\omegak(r_0)} = \left(\frac{r_0}{\rc}\right)^{3/2}\, .  
  \label{eq.fastness}
\end{equation}

The characteristic magnetospheric radius, referred to as the `Alfven radius', is
\begin{equation}
  \ra = \left( \frac{\mu^2}{\dot{M} \sqrt{GM}} \right)^{2/7}.
  \label{eq.ra}
\end{equation}
Here, $\dot{M}$ is the accretion rate, $M$ is the stellar mass, $\mu$ is the stellar magnetic dipole moment.

We introduce a special value of the accretion rate,  $\Mdot_\mathrm{prop}$, called the `propeller accretion rate'. This is the accretion rate at which the Alfven radius equals the corotation radius: $\ra(\Mdotpr)=\rc$. The accretion rate can then be expressed in units of the propeller accretion rate: 
\begin{equation}\label{mdot_small}
    \dot m = \Mdot /\Mdotpr = (\rc/\ra)^{7/2}.
\end{equation}

In the present paper, most results are presented for AMXPs, with some results also shown for two other classes of magnetized astrophysical objects accreting from discs: XRPs and IPs. In Table~\ref{tab:typical}, we list the default parameter values for these types of objects.

\begin{table}
    \centering
    \caption{\label{tab:typical} Assumed parameter values for accreting millisecond pulsars, X-ray pulsars, and intermediate polars. Typical  values of spin frequencies (periods) are: $f_{\rm AMXP} = 100-600$~Hz; $P_{\rm IP} = 10^2-10^3$~s, and  $P_{\rm XRP} = 1-100$~s. }
    \begin{tabular}{|c|c|c|c|c|}
        \hline
        & $\mu$ ~(G\, cm$^{3}$) & $M (M_{\odot})$ & $R_{\rm star}$ (km) & $f$ or $P_{\rm spin}$  \\
        \hline
        AMXP   & $10^{26}-10^{27}$ & $1.4 $ & $11$  & $200$~Hz \\
        \hline
        IP     & $10^{33}-10^{34}$ & $0.7$ & $10^4$ &$10$~min \\\hline
       XRP     & $10^{29}-10^{30}$ & $1.4$ & $11$ &   4 s \\
        \hline
    \end{tabular}
    \tablebib{(1) \citet{Sul19}; (2) \citet{Patr21}; (3) \citet{Tsygankov_review}.}
\end{table}

Using the typical values from Table~\ref{tab:typical}, we can estimate the propeller accretion rates for different objects:

\begin{equation}
\label{eq.Mprop_vals}
    \Mdotpr~[\textrm{g\, s}^{-1}] \approx 
    \begin{cases}
    \displaystyle
2.3\times 10^{15} \,\,
\mu_{26}^2  M_{1.4}^{-5/3}f_{200}^{7/3},&\textrm{AMXP},\\
\displaystyle    
4.7\times 10^{16} \,\,
\mu_{30}^2  M_{1.4}^{-5/3}P_{4 ~\rm s}^{-7/3},&\textrm{XRP},\\
\displaystyle    
1.3\times10^{18}\,\,
\mu_{33}^2 M_{0.7}^{-5/3}P_{10 \rm ~min}^{-7/3},&\textrm{IP}.
    \end{cases}
\end{equation}
We denote the coordinate of the upper surface of the disc by $z_0(r)$, and the disc relative semi-thickness, or aspect ratio, as  $z_0(r)/r \equiv \delta(r)  \ll 1$.

\subsection{Main equations}\label{s.main_eqs}

We consider the case of stationary accretion, $\pd{}{t} = 0$.  Using the equation of mass continuity,
\begin{equation}\label{eq_cont}
    \frac{\partial \rho}{\partial t} + \text{div}(\rho \vb{v}) = 0,
\end{equation}
 we define the accretion rate: 
\begin{equation}
    \Mdot \equiv -\int_{-z_0}^{z_0} 2\pi r \rho v_r \textrm{d}z = \textrm{const} > 0. 
\end{equation}

For the viscous conducting flow of a thin accretion disc in the gravitational and magnetic fields, we adopt the Navier-Stokes equation~\citep[see][]{Kato08}: 
\begin{multline}\label{Nav_Stokes}
  \frac{\partial \vb{v}}{\partial t} + (\vb{v}\vb{\nabla})\vb{v} = -\frac{1}{\rho} \vb{\nabla} P  - \frac{1}{4\pi \rho} \vb{B} \times [\nabla \times \vb{B}]  - \frac{1}{\rho} \vb{\nabla} \Phi_g +\\
  + \frac{1}{\rho} \frac{1}{r^2} \pd{}{r} (r^2 w_{r\fim}) \, \vb{e}_{\fim}. 
\end{multline}
Here $\Phi_g$ is the gravitational potential of the central star:
\begin{equation}
  \Phi_g = -\frac{GM}{\sqrt{r^2 + z^2}},
\end{equation} 
and the last term represents the  viscous force per unit mass. The $\alpha$-prescription for the turbulence model is adopted \citep{SS73}: $w_{ij}=0$ if $ij \neq r\fim$ and we relate the $r\fim$-component of the viscous stress tensor to the pressure via $w_{r\fim} = -\alpha P$, where $P$ is the gas pressure and $\alpha$ is assumed to be constant. 

The three components of the Navier-Stokes equation \eqref{Nav_Stokes} are:
\begin{equation}\label{main}
\begin{cases}
{\displaystyle
(\vb{v}\nabla)v_r - \frac{v_{\fim}^2}{r} = -\frac{1}{\rho} \frac{\partial P}{\partial r} - \frac{GM}{r^2} -  \frac{\vb{\mathcal{M}}_r}{\rho},
}\\
\displaystyle
(\vb{v}\nabla)v_{\fim} + \frac{v_r v_{\fim}}{r} = -\frac{1}{\rho r} \frac{\partial P}{\partial \fim} + \frac{1}{\rho r^2} \frac{\partial}{\partial r} (r^2 w_{r\fim}) - 
 \frac{\vb{\mathcal{M}}_{\fim}}{\rho} ,\\
\displaystyle
(\vb{v}\nabla)v_z = - \frac{1}{\rho}\frac{\partial P}{\partial z}  - \frac{GM}{r^3} z  - \frac{\vb{\mathcal{M}}_z}{\rho}. \\
\end{cases}
\end{equation}
Here  $\vb{\mathcal{M}} = \frac{1}{4\pi}\vb{B} \times [\nabla \times \vb{B}]$ is the Lorentz force. 

The first equation governs the angular velocity in the disc.
The second equation, the equation of angular momentum transport, is solved in Sect.~\ref{sec:radial_str} to determine the radial structure.
The third equation, which describes vertical magnetohydrostatic equilibrium, is used to solve for the vertical structure in Sect.~\ref{sec:vert_str}.

Since  the disc is axisymmetric, these equations can be averaged over the $\fim$-angle. For an arbitrary quantity $x(r, \fim, z)$, we define:
\begin{equation}
    \langle x\rangle  \equiv \frac{1}{2\pi}\int_0^{2\pi} x(r, \fim, z)\, {\rm d}\fim\, .
\end{equation}
This procedure is justified as long as the stellar rotation period, $1/f$, is shorter than the time required for the vertical structure to settle into equilibrium, namely 
the thermal timescale, $t_{\rm th} \sim 1/(\alpha \omegak)$. Consequently,  condition   $1/f < 1/(\alpha \omegak)$  results in 
\begin{equation}\label{eq:cond_aver}
    r/\rc \gtrsim \alpha^{2/3} \sim 0.2 \,\alpha_{0.1}^{2/3}. 
\end{equation}

Alternatively, with Eq.~\eqref{mdot_small} and assuming $r>\ra$,
\begin{equation}
    \dot M  \lesssim 280\, \dot M_{\rm prop} \, \alpha_{0.1}^{-7/3}\, .
\end{equation}
This condition is expected to be satisfied in states that are not very bright, see Eq.~\eqref{eq.Mprop_vals}. In central parts of discs around AMXPs, however, vertical structure of a disc may be azimuth-dependent. 

\section{Interaction between disc and magnetic field}\label{sec:mag_field}

\begin{figure*}
 \includegraphics[width=\textwidth]{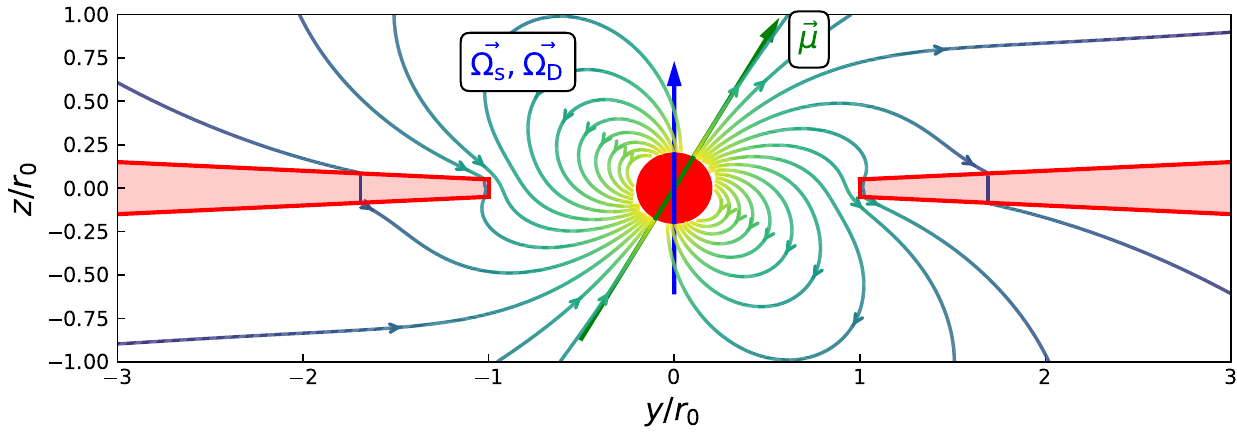}
 \caption{Schematic view of the `star + accretion disc' system in the $(y, z)$ plane (or, equivalently, in the meridional plane $\fim = 90^\circ$). The coordinates are normalized to the inner disc radius, $r_0$. The figure shows the star with $R_{\rm s} = 0.2\, r_0$, the disc with constant relative semi-thickness $\delta = 0.05$, and  magnetic field lines for $\eta = 0.2$ and $\chi = 30^{\circ}$. The field lines are seeded on the stellar surface with a density proportional to the surface field strength. The colour shows the strength of the poloidal field.}
 \label{fig:magn_lines}
 \end{figure*}

The external background field in the absence of the accretion disc is assumed to be purely dipolar within the light cylinder. The full expression is given 
by Eq.~\eqref{dipole_field_full}. In the limit of small heights above the equatorial plane,  $z \ll r$, it reduces to:
\begin{equation}\label{dipole_field}
    \begin{pmatrix}
        B_r\\
        B_\varphi\\
        B_z
    \end{pmatrix}^{\rm dip} = 
      \frac{\mu}{r^3}\,
      \begin{pmatrix}
          2\sin{\chi} \cos{\varphi'} + \mathcal{O}(z/r) \\
          \sin{\chi}\sin{\varphi'} \\
          -\cos{\chi}   +  \mathcal{O}(z/r)
      \end{pmatrix}\, .
\end{equation} 
Here, the angle $\fim'$ in the rotating coordinate system is related to the angle $\fim$ in the inertial reference frame by $\fim' = \fim - \Omega_s t$. 

There is no dipolar field beyond the light cylinder $r_\mathrm{l}$. To account for this,
we assume that $\vb{B} \equiv 0$ for $r > r_\mathrm{out}$. In this paper, when presenting the results, we always assume $r_\mathrm{out}=\rl$, although other possibilities are also discussed. 

Depending on the properties of the disc, different modes of field penetration can be assumed, described in the following subsections. 

\subsection{Diamagnetic disc}
Let us first consider a model of the disc-field interaction under the assumption that the disc is an ideal conductor.
We refer to this model as a `diamagnetic disc'. 

The dipole field of a central star is modified in the presence of the disc in such a way that the magnetic field vanishes inside. For the diamagnetic disc, the three-dimensional solution for the external field was found by \citet{Aly80} \citep[see also][]{Riff80, KR80}. According to this solution, at the surfaces of the razor-thin disc, 
\begin{equation}\label{Aly_pure}
    \begin{cases}
    \displaystyle
        B_r = B_r^{\rm dip} \pm \frac{4}{\pi D} B_z^{\rm dip}\,,\\
    \displaystyle
        B_{\fim} = B_{\fim}^{\rm dip}\,,\\
    \displaystyle
        B_z = 0.
    \end{cases}
\end{equation}
In the first line, the sign $+/-$ corresponds to the upper/lower surface of the disc, and the dipole components are
taken in the equatorial plane, $z = 0$. Here, $D$ is the amplification coefficient\footnote{This amplification resembles the well-known electromagnetic effect near a sharp edge of a conductor. If the disc opening angle satisfies $\delta \ll 1$, the field amplification near the edge is $B^{\rm ampl}/B^{\rm init} \sim \delta^{-1/2}$.  For discussion, see Secs.~3 and 4 of \citet{Lip78} or Sec.~4 of \citet{Aly80}.}
of the magnetic field near the inner edge of the disc and can be written as follows~\citep{Lai99}:
\begin{equation}\label{coef_D}
    D = \textrm{Max}\left(\sqrt{\left(\frac{r}{r_0}\right)^2 - 1},  \, \sqrt{2\delta_0}\right),
\end{equation}
where $\delta_0 = z_0(r_0) / r_0$ is the relative half-thickness at the inner edge of the disc.

\subsection{Magnetically threaded disc}\label{s.conductivedisc}

The second model, or the `conductive' disc, or the `finite-conductivity' disc, assumes that the NS dipole field penetrates the disc because of its non-zero resistivity, provided that sufficient time has elapsed for this to occur~\citep{Livio-Pringle1992, Wang1996, KR07}. The effective resistivity is due to magnetic diffusivity and is thought to originate from disc turbulence, which also underlies the disc viscosity. Stationary axisymmetric components of the field penetrate the disc, see \citet{Lai99}, because the opposing currents have decayed.
The magnetic field inside the disc is then a superposition of the stationary dipole field and the toroidal field induced by the difference between the angular velocities of the magnetosphere and the disc.\footnote{Notice that, in contrast to the model of \citet{Lai99}, the rotation axes of the disc and star are aligned in our model, and the leading term in $B_z^{\rm dip}$ is constant; see Eq.~\eqref{dipole_field}.}

\citet{Wang97} proposed the following expressions for the field induced in the disc in the case where turbulent diffusion is the mechanism responsible for magnetic field dissipation:
\begin{equation}\label{param_II}
b_{\fim}^\mathrm{W}
= 
 \begin{cases}
 \displaystyle
   -\Gamma \left(1 - \frac{\omegas}{\omegak(r)}
   \right) B_z^{\rm dip}(r, \fim, z=0) &\text{surface},\, \emph{z} = \emph{z}_0,\\
\displaystyle
   \Gamma \left(1 - \frac{\omegas}{\omegak(r_0)}\right) B_r^{\rm dip}(r_0, \fim, z=0) &\text{inner edge}, \,\emph{r} = \emph{r}_0,\\
\displaystyle
   \textrm{not specified}, &\text{inside the disc}.
 \end{cases}    
\end{equation}

The coefficient $\Gamma$ is an unknown dynamo coefficient of order of unity.

Alternatively, under the same assumption, by solving the induction equation, \citeauthor{Kuz23} (\citeyear{Kuz23}, hereafter Paper I) derived the form of the induced field $b_{\fim}(r, z)$. Since both $B_z^{\rm dip}$ and $B_r^{\rm dip}$ depend on $\fim'$ as $C_1 + C_2 \cos{\fim'}$ (see Eq.~\eqref{dipole_field_full}), the general form of the induced field is
\begin{equation}
    b_{\fim} \equiv b(r, z) + b_1(r, z) \cos{\fim'}\, ,
    \label{bphi_small}
\end{equation}
where the first term is the axisymmetric stationary component and the second term is the non-axisymmetric, non-stationary component (pulsating, since $\fim' = \fim - \omegas t$).

The induced magnetic field was found to depend on the inner radius of the viscous disc, $r_0$, and on the relative half-thickness of the disc, $z_0(r)/r$. The induced field was also required to vanish beyond the radius $r_{\rm out}>r_0$, where the stellar field lines disconnect from the disc (the author assumed this to occur at the light cylinder radius $r_\mathrm{l} = c/\omegas$).
 
\begin{figure}[ht!]
\centering
 \includegraphics[width=\hsize]{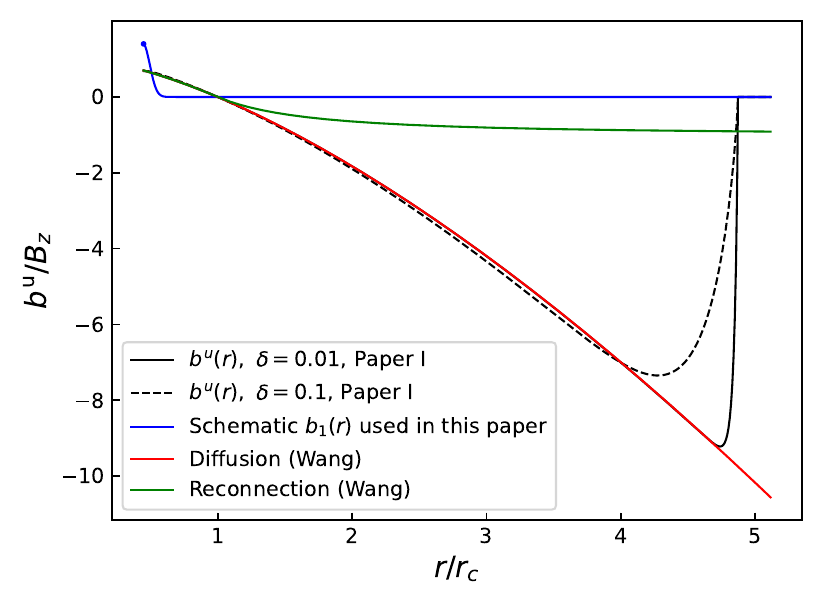}
 \caption{Radial dependencies of $b^\mathrm{u}(r)$ from \citet{Kuz23} (for two different relative half-thicknesses $\delta$), together with Wang's `diffusion' and `reconnection' models. The schematic dependence of $b_1(r)$ is also shown, illustrating the sharp decline from $b_1(r_0)$ to $0$ near $r_0$. The inclination angle is $\chi = 45^{\circ}$.}
 \label{fig:magn}
\end{figure}

 According to Paper I, 
 the stationary axisymmetric part of the induced field depends almost linearly on $z$:
 \begin{equation}
     b(r, z) \approx b^\mathrm{u}(r) \frac{z}{z_0} \, ,
     \label{eq.brz}
 \end{equation}
 where the field at the upper surface of the disc, $b^\mathrm{u}(r)$, closely resembles Wang's `upper surface' component with $\Gamma=1$. The numerical solutions for $b^\mathrm{u}(r)$ from Paper I, obtained for different disc thicknesses, are compared with Wang's analytic prescription for $\Gamma=1$ in Fig.~\ref{fig:magn}.
Since the numerical solution requires significant computational resources while providing the results virtually identical to the analytical approach, we use the latter in this work.

In addition to the stationary axisymmetric component \eqref{eq.brz}, in the vicinity of the inner radius we also consider, following \citet{Wang97}, the non-axisymmetric oscillating component of the induced field corresponding to the second term in Eq.~\eqref{bphi_small}. This component is absent for an aligned dipole, but otherwise arises because of the twisting of magnetic field lines near the inner radius. Thus, approximately, it is proportional to the radial component of the dipole field (see Eq.~\eqref{dipole_field}) and independent of $z$. This component
vanishes within the disc as the radial distance from the inner edge increases beyond the characteristic `skin effect' length scale (see Fig.~\ref{fig:magn}):
\begin{equation}\label{b1_r}
    b_1(r) = \frac{2\mu \sin{\chi}}{r_0^3}\left(1 - \frac{\omegas}{\omegak(r_0)}\right) \, \times \exp\left(-\frac{r-r_0}{d_{\rm skin}}\right)\, .
\end{equation}
Hereafter, we assume $d_{\rm skin} \approx z_0$, see also Appendix~\ref{App:MF}. 

Summing up, we write the induced magnetic field in the model of the conductive disc as in Eq.~\eqref{bphi_small},
where Eqs.~\eqref{eq.brz}~and~\eqref{b1_r} are to be substituted together with the following expression for the magnetic field at the disc surface:  

 \begin{equation}\label{eq.bur}
    b^\mathrm{u}(r) = 
    \begin{cases}
    \displaystyle
    \frac{\mu}{r^3}\left(1-\frac{\omegas}{\omegak(r)}\right) \, \cos\chi   \,, & r_0 \leq r \leq r_\mathrm{out}, \\
    0, & \mathrm{otherwise.}
    \end{cases}
 \end{equation}

The angular velocity of matter in the disc is not strictly Keplerian. However, \citet{Campbell92, Campbell09} argued that, in thin discs, deviations from Keplerian rotation become significant only near the magnetospheric boundary layer, where the magnetic pressure gradient is large. This issue is studied in Appendix~\ref{subsec:omega}, where the main conclusion is that the Keplerian approximation, $\Omega = \omegak$, remains valid provided that the magnetic field is not too strong.

\subsection{Intermediate case} \label{s.intermediate}
A `hybrid' model between the diamagnetic and magnetically threaded disc models was proposed by \citet{Lai99} for the case of misaligned rotation axes of the disc and star. The idea is that the time-dependent $z$-component of $\vb{B}^{\rm dip}$ does not penetrate the disc, whereas the stationary (with respect to the disc) component of the dipole field, denoted $B_z^{\rm in}$, does. The screened part of the field corresponds to a time-dependent surface current, $|K_\varphi| \propto B^{\rm dip}_z - B^{\rm in}_z$, which generates an additional external radial component of the magnetic field, $B_r$.

In a similar way, we bridge the case of an ideally conducting disc and the case of finite conductivity. In our configuration, where the rotation axes of the disc and star are aligned, only the $z$-component of the magnetic field has a stationary leading term, see Eq.~\eqref{dipole_field}. We propose that only a fraction of the stationary $B_z$ effectively penetrates the disc: $B_z^{\rm in} = \eta B_z^{\rm dip}$, with $0\leq\eta\leq1$. The toroidal field induced by the rotation of the disc is then $\eta b_{\fim}$. The part of the field that does not penetrate the disc follows the Aly solution:
\begin{equation}
B_r = B_r^{\rm dip} + \frac{4}{\pi D} (B_z^{\rm dip} - B_z^{\rm in})\, 
\label{eq.Lai}
\end{equation}
and the coefficient $D$ describes the field amplification near the inner edge of the disc; see Eq.~\eqref{coef_D}.

The advantage of this parametrization is that it naturally reproduces the limiting cases of the diamagnetic disc at $\eta = 0$ (no field penetrates the disc surface, corresponding to the Aly solution) and the threaded disc at $\eta = 1$ (complete penetration of the field into the disc, corresponding to the Paper I or Wang model). We show\footnote{Magnetic field lines are shown as streamlines computed with the \texttt{streamplot} routine from Matplotlib v. 3.10.8 \citep{Hunter07}.} the poloidal magnetic field for the case $\eta = 0.2$ and inclination $\chi = 30^{\circ}$ in Fig.~\ref{fig:magn_lines}.

In Appendix~\ref{App:MF}, we provide the complete formulas for the magnetic field inside and outside the disc. In this model, the field is discontinuous at the disc surfaces. The azimuth-averaged discontinuities of the magnetic pressure, $\Delta_z P_{\rm m} \equiv P_{\rm m}(z_0^+) - P_{\rm m}(z_0^-)$ and $\Delta_r P_{\rm m} \equiv P_{\rm m}(r_0^-) - P_{\rm m}(r_0^+)$, are easily calculated:
\begin{equation}\label{delta_Pmagn}
    \Delta_z P_{\rm m} 
    = \frac{\mu^2}{8\pi r^6} \left(\frac{5}{2} \sin^2{\chi} + \left(\frac{4(1-\eta)}{\pi D}\right)^2 \cos^2{\chi}\right)\, ,   
\end{equation}
\begin{equation}
    \Delta_r P_{\rm m} 
    = \frac{\mu^2}{8\pi r_0^6} \left(\frac{1}{2} \sin^2{\chi} + \frac{8}{\pi}\left(\frac{(1-\eta)^2}{\pi \delta_0}+\frac{\eta(1-\eta)}{\sqrt{2\delta_0}}\right)\cos^2{\chi}\right)\, .   
\end{equation}
The latter expression may be rewritten as
\begin{equation}\label{delta_Pmagn_r}
    \Delta_r P_{\rm m} 
    \equiv \frac{\mu^2}{\pi^3 r_0^6 \delta_0} f_P(\eta, \chi, \delta_0),  
\end{equation}
where the function $f_P(\eta, \delta_0, \chi)$ is defined as
\begin{equation}\label{fp}
f_P = \left((1-\eta)^2 + \eta~(1-\eta) ~\pi \sqrt{\frac{\delta_0}{2}}\right)\cos^2{\chi}  + \frac{\pi^2\delta_0}{16} \sin^2{\chi} . 
\end{equation}

The value of $f_P$ takes a small non-negative value proportional to $\sin^2\chi$ for $\eta\approx 1$ and increases monotonically with decreasing $\eta$, reaching unity for $\eta=0$ and $\chi=0^\circ$. Equation~\eqref{delta_Pmagn_r} reduces to the result of \citet{Aly80} for the magnetic pressure at the inner boundary in the case of a diamagnetic disc, $\eta=0$.

\section{Disc structure}\label{sec:method}

In this section, we first consider the angular momentum transport equation, from which follows the position of the inner disc radius, $r_0$, as well as the viscous stress tensor $\Wrf(r)$ can be determined. We then examine the equations governing the vertical structure.

\begin{figure}
 \includegraphics[width=\columnwidth]{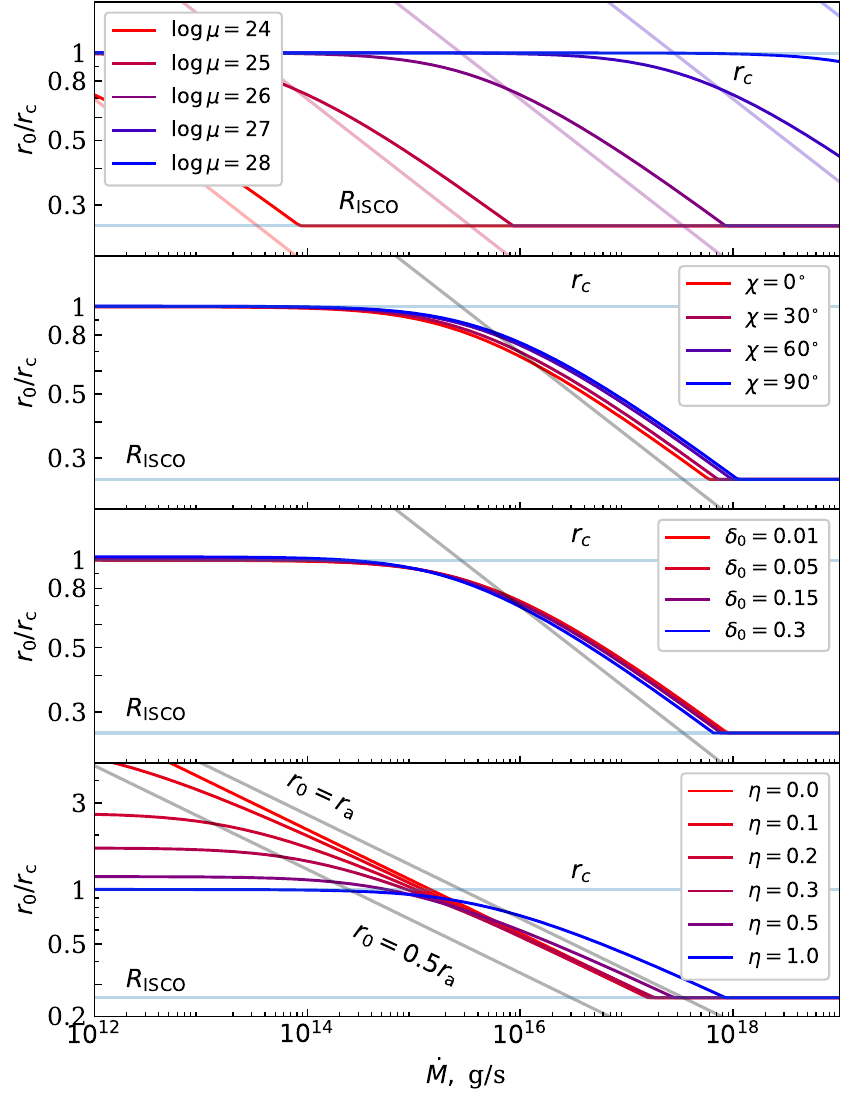}
 \caption{ Dependence of the inner radius of a viscous accretion disc (in units of the corotation radius) on the accretion rate $\dot{M}$. Fiducial parameters are those of an AMXP, with $\chi = 45^{\circ}$, $\delta_0 = 0.05$, $\varkappa = 0.02$, $\eta = 1$. For each value of $\mu$, the relation $r_0 = \ra$ is shown, except in the last panel, where both $r_0=\ra$ and $r_0 = 0.5 \, \ra$ are displayed.}
 \label{fig:r0}
\end{figure}

\begin{figure}
 \includegraphics[width=\columnwidth]{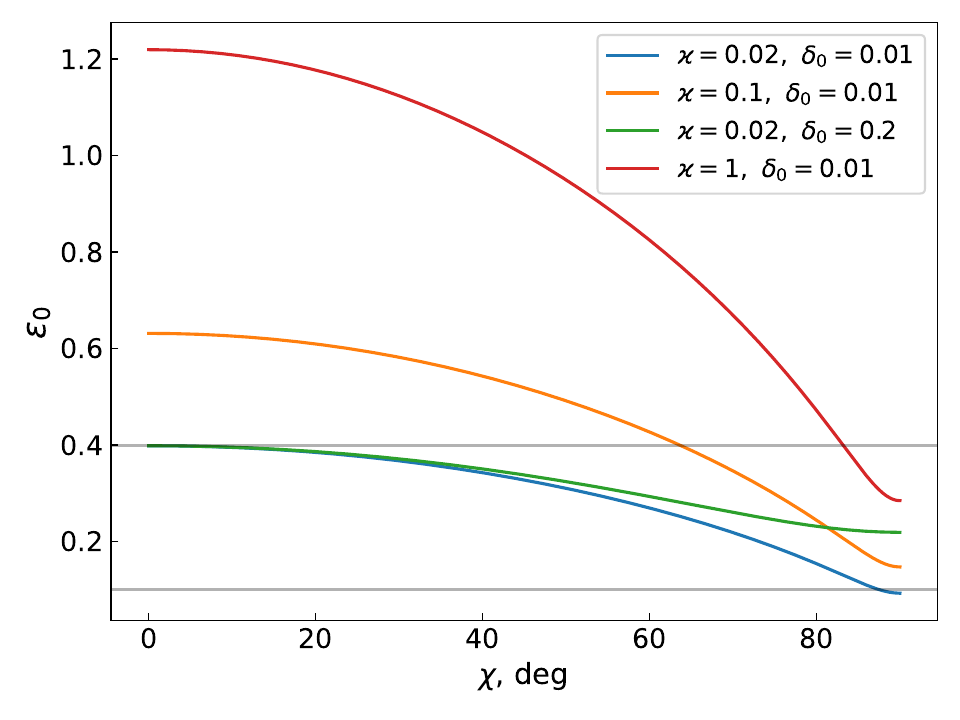}
 \caption{Dependence of $\varepsilon_0(\eta=0, \chi, \delta_0)$ equal to $r_0/\ra$ (see Eq.~\eqref{r0_eta0}) on the magnetic inclination angle $\chi$. The disc aspect ratio $\delta_0$ at the inner radius and the parameter $\varkappa \sim {2\alpha}/{\pi^2}$ are indicated in the legend. Horizontal lines are drawn at $\varepsilon_0=0.1$ and $\varepsilon_0=0.4$.} \label{fig:eps0}
\end{figure}

\subsection{Radial structure}\label{sec:radial_str}
To find the viscous torque in the disc, we average the second equation of system \eqref{main} over $\fim \in [0, 2\pi]$ and integrate it over $z \in [-z_0, z_0]$.
Denoting the vertically integrated stress tensor $W(r) = \int_{-z_0}^{z_0} w(r, z) \dif z$, we obtain (see Appendix~\ref{sec:ang_mom} for details):
\begin{multline}\label{Wrphi_new}
    \Wrf(r) =  \left(\frac{r_0}{r}\right)^2 \Wrf(r_0) - \frac{\dot{M} \Omega_k}{2\pi} \left(1 - \sqrt{\frac{r_0}{r}}\right) +\\ +
\frac{1}{2\pi r^2}\,\int_{r_0}^{r}  \frac{\dif F_{\rm magn}}{\dif r} \dif r  
    \, 
\end{multline}
 and the viscous heat flux per the unit area emitted from one surface of the disc is
\begin{equation}\label{eq.Q0}
    Q_0(r) = \frac{3}{4} \, \omegak \, |\Wrf|.
\end{equation}
Here, according to the $\alpha$-parametrization of viscosity, the viscous stress tensor  at the inner boundary is proportional to the magnetic pressure discontinuity $\Delta_r P_{\rm m}$, see \S\ref{s.intermediate}:
\begin{equation}\label{New_cond_W}
   \Wrf(r_0) = -\frac{\mu^2}{r_0^5}\,\frac{2\alpha}{\pi^3}f_P(\eta, \chi, \delta_0)\, .
\end{equation}

On the right hand side of Eq.~\eqref{Wrphi_new}, the first term corresponds to a disc-reservoir, or a `dead disc'~\citep{SS77}, a solution with zero accretion rate and a non-zero torque at the inner edge; the second term is the standard-disc expression, and the third term arises from the magnetic torques integrated over radius:
\begin{equation}\label{Fmagn}
    \frac{\dif F_{\rm magn}}{\dif r} = - \frac{\mu \eta^2 \cos{\chi}}{r}b^\mathrm{u} 
     + \mu \eta^2 \sin{\chi} \frac{\dif}{r^{3/2}\dif r}\left(b_1 \frac{z_0}{r} r^{3/2}\right).
\end{equation}

The last expression explicitly shows that the magnetic torque depends on the relative thickness of the disc. Since $b_1$ is localized near $r_0$, the quantity $z_0/r$ should be taken at the inner radius. In view of the potentially very complex physical processes occurring at the disc inner edge~\citep[including instabilities such as Kelvin--Helmholtz or interchange instabilities, disc fragmentation, MHD reconnection, etc.;][]{AronsLea76, Lisitsin-Shakura2023}, which lie beyond the scope of our model, we treat the relative half-thickness at $r_0$ as a free parameter and denote it by $\delta_0 \equiv z_0(r_0)/r_0$. Since the structure of the disc does not appear to be sensitive to the exact value of $\delta_0$ (see below), as long as it remains $\lesssim 0.1-0.2$, we henceforth assume $\delta_0 = 10^{-2}$, unless stated otherwise.

To calculate $\Wrf$ analytically, we adopt the form of the induced field proposed earlier in  Eq.~\eqref{bphi_small}. Evaluating the integral in Eq.~\eqref{Wrphi_new} using the induced magnetic field \eqref{eq.bur}, and taking into account that $b_1$ is negligible far from the inner radius, i.e. for $r-r_0 \gg r_0\delta_0$ (see Eq.~\eqref{b1_r}), we obtain:
\begin{equation}
\label{eq.wrf_wang}
     \Wrf (r) = -\frac{\dot{M} \omegak}{2\pi} 
     \,\left(1 - \sqrt{\frac{r_0}{r}}\right) - \frac{\mu^2}{2\pi r^2 r_0^3} g(r/r_0)\, .
\end{equation} 

 The first term 
 represents the standard-disc solution with inner radius $r_0$. The second term therefore describes the modification of the standard stress tensor caused by the presence of the magnetic field. 
The absolute value of the dimensionless $g(r/r_0)$ is either constant or starts from a small value at the disc's inner edge and reaches an asymptotic constant value of $-0.2\dots0.4$, within a few $r_0$ (see Appendix~\ref{subsec:W}). For $\eta = 0$ (the Aly solution), even though $F_\mathrm{magn}=0$, the viscous stress at the inner radius is non-zero, so $g \neq 0$. The viscous stress $\Wrf(r)$ calculated according to Eq.~\eqref{eq.wrf_wang} is always negative and vanishes at the inner disc radius when $\eta=1$ and $\chi=0$, in agreement with the conjecture of \citet{KR07}.

To evaluate the viscous stress tensor, the formula above requires the inner disc radius $r_0$ to be known.

In the literature, the inner radius of the disc is defined differently in the diamagnetic and threaded disc models. In the former, the inner radius is commonly defined as the distance from the star at which the magnetic pressure balances the disc pressure. In this case, the inner radius of the disc is expected to be comparable to the Alfven radius~\eqref{eq.ra}: $r_0\sim \ra$. In threaded-disc models, the inner radius is determined from a condition imposed on the viscous torque at the inner radius. For example, \citet{Wang95, KR07} use the vanishing torque condition.

We propose the following bridging formula for the inner radius $r_0$ of the disc partially shielding the stationary magnetic field:
\begin{equation}\label{eq.r0_new}
        r_0 =  \ra  \times \left(
        \varepsilon_0 ^{7/2} + 2\,\eta^2 \,(1-\omega)\, {f_\chi}\right)^{2/7}\, ,
\end{equation}
where $\varepsilon_0 \equiv (2\varkappa f_P)^{2/7}$, with $\varkappa \sim 2\alpha/\pi^2$, and $f_P$ is defined by Eq.~\eqref{fp}. Thus, $\varepsilon_0$ may depend on the turbulent parameter $\alpha$, the disc half-thickness at the inner radius $\delta_0$, the magnetic inclination angle $\chi$, and the penetration parameter $\eta$, and is expected to satisfy $\varepsilon_0 < 1$ if $\varkappa<0.5$. {The `inclination factor' $f_\chi \equiv 1 + (1-3\delta_0)\sin^2\chi$ collects the contributions of the axisymmetric ($\propto\cos^2\chi$) and non-axisymmetric ($\propto\sin^2\chi$) induced fields at the inner edge.} We give the derivation of this equation in Appendix~\ref{sec:ang_mom}.
Further details regarding the definition of $\varkappa$ are given after Eq.~\eqref{eq_r0_eta_app}.

We note that, since the fastness $\omega$ depends on $r_0$ (cf.~definition \eqref{eq.fastness}), the equation above is transcendental with respect to $r_0$. If we ignore the magnetic field discontinuity at the inner edge of the disc (see Sect.~\ref{s.intermediate}), then $\varepsilon_0=0$ and the radius \eqref{eq.r0_new} reduces to the expression very similar to those proposed by \citet{Wang97, bozzo} (identical, with $\chi=0$ and $\Gamma=1$). 

In Fig.~\ref{fig:r0}, we explore how the inner radius behaves for varying magnetic moments $\mu$, inclination angles $\chi$, inner-edge half-thicknesses $\delta_0$, and penetration parameters $\eta$. 

For the case of $\eta \neq 0$, the behaviour of the inner radius $r_0$ differs 
between two regimes:  `strong' and `weak' accretion.  We define the `strong' (`weak') accretion regime by $\ra < \rc$ ($\ra > \rc$), or equivalently by $\dot M > \dot M_{\rm prop}$ ($\dot M < \dot M_{\rm prop}$).
One can see from Eq.~\eqref{eq.r0_new}
that in the  case of `strong' accretion, when the disc intrudes well under the corotation radius,  the inner radius is proportional to the Alfven radius:
\begin{equation}\label{r0_strong}
    r_0 \approx \ra \times  \varepsilon_0 \left(1+\frac{2\,\eta^2 {f_\chi}}{\varepsilon_0^{7/2}} \right)^{2/7}.
\end{equation}

In the `weak' accretion regime, the inner radius reaches a maximum value of
\begin{equation}\label{r0_weak}
    r_0 = \rc \, \omega_{\rm max}^{2/3}\, ,    
\end{equation}
where $\omega_{\rm max}$ is found by taking the limit $r_0/\ra \rightarrow 0 $:
\begin{equation}\label{crit_w_new}
   \omega_{\rm max} = 1 + \frac{\varepsilon_0^{7/2}}{2\eta^2\,{f_\chi}}\, .
\end{equation}

Therefore, for a magnetically threaded disc, the growth of the inner radius is limited by a maximum value close to the corotation radius. This departs from the assumption that the inner radius is proportional to the Alfven radius~\citep[see also][]{KR07}. The maximum value of $r_0$ depends on the magnetic inclination angle $\chi$~\citep[see also][]{Wang97,bozzo} and only weakly on $\delta_0$. {Figure~\ref{fig:r0comp} compares $r_0(\Mdot)$ from Eq.~\eqref{eq.r0_new} for different $\eta$ with earlier prescriptions.}

In the case of a diamagnetic disc, $\eta=0$, and constant parameter $\varkappa$, the inner radius is always proportional to the Alfven radius: 
\begin{equation}\label{r0_eta0}
    r_0 = \ra \, \left[2\varkappa f_P(\eta=0, \chi, \delta_0)\right]^{2/7}  =  \ra \times \varepsilon_0(\eta=0) \,.
\end{equation} 

We show how $\varepsilon_0(\eta=0)$ depends on the inclination angle $\chi$ in Fig.~\ref{fig:eps0}. For example, if $\varkappa = 0.1$, then $r_0 \approx 0.63 \, \ra$ for $\chi=0^\circ$ and $r_0 \approx 0.32\, \ra$ for $\chi=80^\circ$. For inclination $\chi=45^\circ$ and $\varkappa=1$, the inner radius $r_0 \approx \ra$. In the remainder of the paper, we adopt $\varkappa = 0.02$.

\subsection{Vertical structure}\label{sec:vert_str}
To calculate the vertical structure, we modified the code \texttt{AlphaDiSC}, first presented in TLM23. In this section, we provide a brief overview of the method used in TLM23, with particular emphasis on aspects related to the presence of the magnetic field.

Averaging the equation of the vertical hydromagnetic equilibrium (the last Eq. in \eqref{main}) over $\fim \in [0, 2\pi]$, and assuming $v_z=0$, we obtain
\begin{equation}\label{vert_first}
    \pd{P}{z} = -\rho \Omega^2_{\rm K} z - \frac{1}{4\pi} \langle \left(\vb{B} \times [\nabla \times (b_{\fim} \vb{e}_{\fim})]\right)_z \rangle_\fim \, ,
\end{equation}
where $\bm{B}$ is the total magnetic field in the disc, given by Eq.~\eqref{eq.B_eta_full}, and $b_{\fim}$ is the induced toroidal component defined in Eq.~\eqref{bphi_small}. We assume below that the azimuth-dependent component $b_1$ of the induced field is absent (this assumption breaks down only near the inner edge of the disc). Consequently, the equations of hydromagnetic equilibrium, energy transport, and energy generation can be written as follows:
\begin{equation}\label{vert_1}
 \begin{cases}
\displaystyle
       \frac{\dif (P + b^2/8\pi)}{\dif z} =  -\rho\, \Omega_k^2\, z , \\
\displaystyle
    \frac{\dif  \ln{T}}{\dif  \ln{P}} \equiv \nabla = 
    \begin{cases}
        \nabla_{\rm rad}, \nabla_{\rm rad} \leq \nabla_{\rm ad},\\
        \nabla_{\rm conv}, \nabla_{\rm rad} \geq \nabla_{\rm ad},\\
    \end{cases}\\
\displaystyle
   \frac{\dif Q}{\dif z} = \frac{3}{2} \Omega_{k} \alpha P\, , \\
 \end{cases}    
\end{equation}
where $Q$ is the radiative flux along the $z$-axis, and $\nabla$, $\nabla_{\rm conv}$, etc. are the temperature gradients used to account for convection. In the case $b \equiv 0$, the equations reduce to those discussed in TLM23. The code \texttt{AlphaDiSC} employs analytic or tabulated equations of state and opacities from MESA \citep{Paxton11}. All results in this work are presented for solar metallicity and radiative--convective energy transport (method \texttt{structure='MesaRadConv'} in the code). We neglect radiation pressure.

To explore how the addition of $b^2/8\pi$ affects the behaviour of the pressure profile, one may use the approximation \eqref{eq.brz}: $b \approx b^\mathrm{u} \, z/z_0$. Taking the derivative $\dif (b^2/8\pi)/\dif z$ explicitly and moving this term to the right-hand side, we find that the first equation in \eqref{vert_1} may be written as
\begin{equation}\label{eq:m}
    \frac{\dif P}{\dif z} = -\rho \omegak^2 z\, (1 + m),
\end{equation}
with $m = (b^\mathrm{u})^2/(4\pi z_0^2 \rho \omegak^2)$. In the central plane of the disc, $\rho = \rho_{\rm c}$, we obtain 
$m \sim m_c$, where $m_c \equiv \eta^2 \alpha \delta \, (\ra/r)^{7/2}$. Thus, the magnetic contribution satisfies $m_c \ll 1$ always for a diamagnetic disc.

For a magnetically threaded disc with $\eta \sim 1$, however, it is possible that $r_0 < \ra$; see Fig.~\ref{fig:r0}. This corresponds to the weak-accretion regime (see Eq.~\eqref{r0_weak}), in which the inner radius never becomes smaller than the corotation radius. Substituting $\eta=1$ and $r = \rc$ into $m_c$, we find that the magnetic contribution may become substantial at low accretion rates; see Eq.~\eqref{eq.m_magn}.

Following \citet{KR07}, we do not include Joule heating as an additional source of energy generation. 
Inside the disc,  Joule heat may be comparable to the viscous heat, but only very close to $r_0$
(see Appendix~\ref{sec.joule}).
It can additionally hinder recombination of the disc material (see Sect.~\ref{sec:S-curves} below). However, the estimates that can be made within the framework of our model near the inner radius are of limited value. Furthermore, close to $r_0$, the assumptions of a Keplerian accretion flow and a rigidly rotating magnetosphere
become increasingly questionable. Thus, among the equations of system \eqref{vert_1}, only the hydrostatic-equilibrium equation is affected by the presence of the magnetic field. 

System \eqref{vert_1} has a single free parameter, $z_0$, the disc semi-thickness. If there is a magnetic-pressure difference $\Delta_z P_{\rm m}$ between the regions above and below the disc surface, as defined by Eq.~\eqref{delta_Pmagn}, it compresses the disc along the $z$-direction. The boundary conditions at the disc surfaces are imposed as described in TLM23, with the additional magnetic compression term $\Delta_z P_{\rm m}$ added to the surface pressure ($\Delta_z P_{\rm m}$ is added to the right-hand side of Eq.~13 in TLM23). The boundary conditions also depend on $z_0$. The code \texttt{AlphaDiSC} solves system \eqref{vert_1} using a shooting method, varying $z_0$ in order to satisfy the boundary conditions.

\begin{figure}
 \includegraphics[width=\columnwidth]{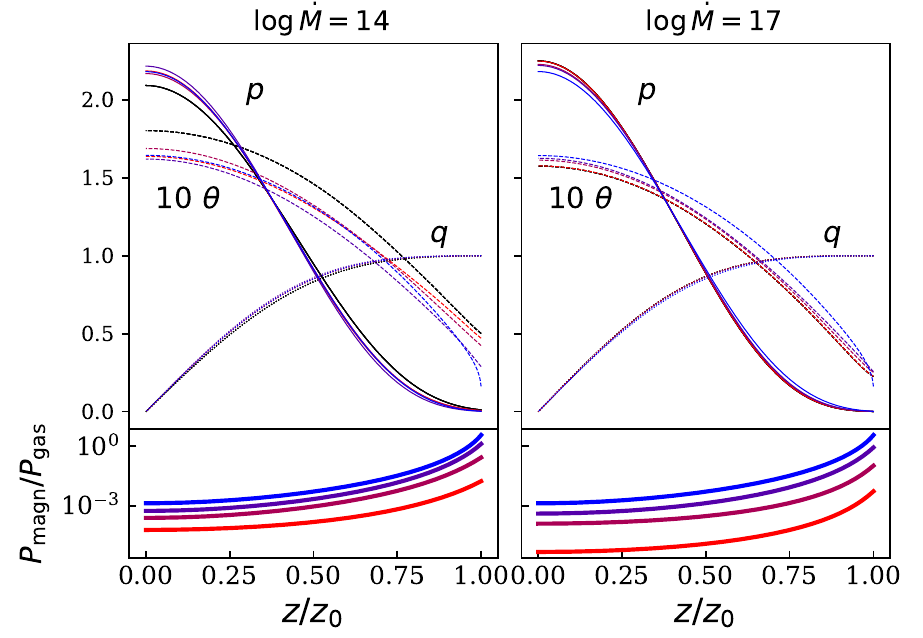}
 \caption{Dimensionless pressure $p = P/P_0$, temperature $\theta = T/T_0$, and flux $q = Q/Q_0$ as functions of the vertical coordinate for the `weak-accretion' (left) and  `strong-accretion' (right) regimes. The ratio of magnetic pressure to gas pressure is shown in the narrow panels. The typical parameters for an AMXP are adopted, with $\chi = 30^{\circ}$. The structure is calculated at $r = 2\,r_0$. The black curves correspond to the standard disc case.}
\label{fig:vert_struct}
\end{figure}

\section{Results}\label{sec:results}

\subsection{Vertical profiles}\label{subsec:vert_results}

Following TLM23, we normalize the pressure and temperature to:
\begin{equation}\label{normal}
    \begin{cases}
    \displaystyle
        P_0 = \frac{2}{3} \frac{Q_0}{ \alpha \, \omegak z_0},\\
        \displaystyle
        T_0 = \frac{\mu \, z_0^2 \, \omegak^2}{\mathcal{R}}.
    \end{cases}
\end{equation}
where ${Q_0}$ is proportional to $\Wrf$, see Eq.~\eqref{eq.Q0}.
Examples of profiles of dimensionless pressure $p = P/P_0$, temperature $\theta = T/T_0$ and energy flux $q = Q/Q_0$ are shown in Fig.~\ref{fig:vert_struct}. 

The vertical profiles of the dimensionless pressure $p$ and flux $q$ do not change significantly compared to the non-magnetic case. This is due to the adopted normalization quantities, which themselves are modified by the magnetic field. The temperature profile behaves differently for the same reason. This result remains unchanged for other astrophysically relevant parameter sets, including those corresponding to IPs and XRPs.

\subsection{S-curves}\label{sec:S-curves}
The thermal stability of accretion discs plays an important role in interpreting the behaviour of transient and persistent binary sources. To analyse the thermal stability of discs, it is useful to consider the so-called S-curves. Let us define the disc surface density as
\begin{equation}
    \Sc = \int_{-z_0}^{z_0} \rho \textrm{d}z\, .
\end{equation}

The S-curve is the dependence of either $\Tc$, $\Teff$, or $\Mdot$ on $\Sc$ at a fixed radius. The concept of S-curves is an important ingredient of the Disc Instability Model \citep[DIM; see][for a review]{Hameury20}. DIM has been successfully applied to explain the observed light curves of dwarf and X-ray novae. Analysing the effects of the external magnetic field on S-curves is therefore an important step towards understanding the evolution of binaries with magnetized components.

\begin{figure}
\includegraphics[width=\columnwidth]{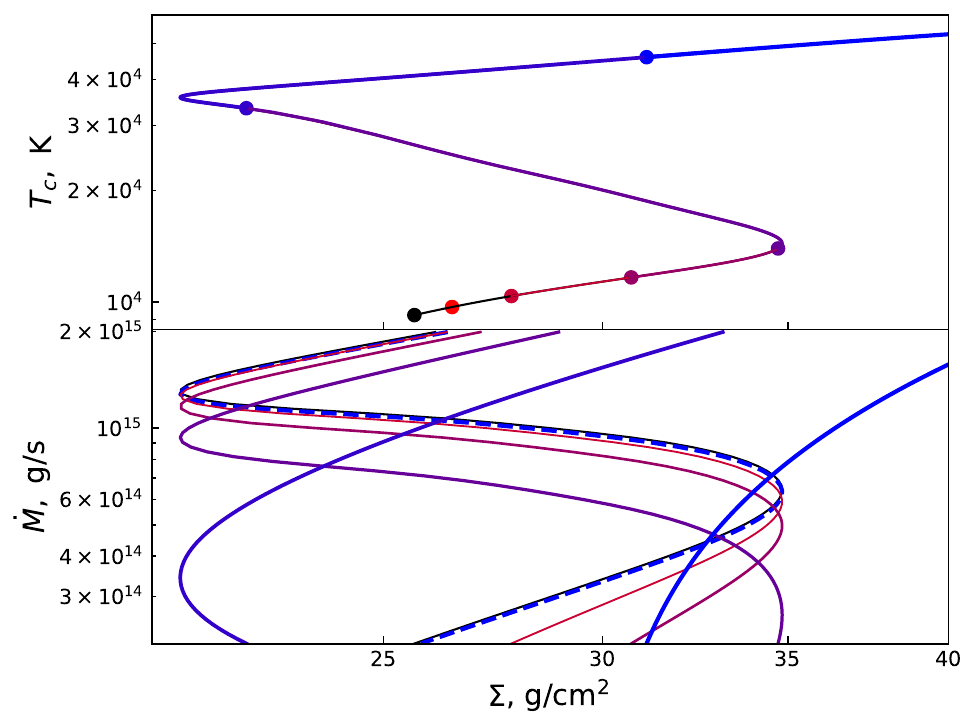}
 \caption{S-curves for an AMXP at distance $\log{r} = 9.7$ from the star. The black colour denotes the standard disc model, while the colours from red to blue represent a grid of surface magnetic fields logarithmically spaced between $\log{B}=8$ and $\log{B}=9$. The dashed blue line corresponds to the case $\log{B}=9$ and $\eta=0$. The points in the $T_{\rm c}-\Sc$ plane indicate the lowest temperature attainable at the minimum $\dot M$ shown, depending on $B$. The inclination angle is $\chi=0^\circ$.}
 \label{fig:S_curve}
\end{figure}

Using the \alphadisc code modified as described in Sect.~\ref{sec:vert_str},  we computed several representative S-curves by varying the accretion rate at a fixed radius (see Fig.~\ref{fig:S_curve}). 

We find that, on the $T_{\rm c}$---$\Sc$ and $T_{\rm eff}$---$\Sc$ planes, neither the shape of the curve nor the positions of the turning points depend significantly on the magnetic field\footnote{The relative deviations of the turning points from those of the standard disc are numerical \citep[$\lesssim 2\%$ for $\Sc$ and $\lesssim 0.1\%$ for $T_{\rm c}$ from those in] []{Tavl23}.}. This is not surprising, as both the shape of the S-curve and the positions of the turning points are determined primarily by local properties, such as temperature and opacity, rather than by the radial structure of the disc. Thus, the position of a turning point in these coordinates may be fully determined by its `non-magnetic' value.

However, the S-curves for discs around magnetized objects differ significantly from those of the standard disc in the $\dot{M}$---$\Sc$ plane, particularly for $\eta \sim 1$; see the bottom panel of Fig.~\ref{fig:S_curve}. Clearly, the critical values of the accretion rate, denoted $\Mdot^\pm_{\rm magn}$, depend on the magnetic-field strength and configuration.

From the relation between the effective temperature and the viscous torque, $T_{\rm eff}^4 \propto F_{\rm vis}\,  r^{-7/2}$~\citep[see Eqs.~\eqref{eq.Q0} and \eqref{F_vis}, and eq. 9 in][]{Blaes+2025}, we obtain the following critical accretion rates for accretion discs around magnetized stars:
\begin{equation}\label{eq.Mdot_crit_mag}
  \Mdot^\pm_{\rm magn} \, \left(1 - \sqrt{\frac{r_0}{r}}\right)  =\Mdot^\pm_{\rm non-magn}
  - \dot M_{\rm prop } \,  \sqrt{\frac{r_0}{r}} \, g(r/r_0)\, \omega^{-7/3}\, , 
\end{equation}
where the function $g$ is given by Eq.~\eqref{g_wang}, and the propeller accretion rate $\dot M_{\rm prop } $ is defined in Sect.~\ref{sec:notation}. The critical values for the standard disc, $\Mdot^\pm_{\rm Non-magn}$, obtained in TLM23 are, neglecting the very weak dependence on $\alpha$:
\begin{equation}\label{Mdot_tavl}
    \begin{pmatrix}
        \dot{M}^{+}_{\rm non-magn}\\
        \dot{M}^{-}_{\rm non-magn}
    \end{pmatrix} \approx 
    \begin{pmatrix}
        1.02\\
        0.51
    \end{pmatrix} \cdot 10^{16}~\textrm{g~s}^{-1}~ M_1^{-0.84} r_{10}^{2.6}.
\end{equation}

Depending on the sign of $g(r/r_0)$ (see Fig.~\ref{fig:g_vs_x}), the critical values of $\dot M$ for a disc subject to magnetic torques can be either lower or higher than the standard values. The difference between the critical rates becomes larger as $\eta\rightarrow 1$, whereas for $\eta=0$ the modification of $\Mdot^\pm$ arises mainly from the modified inner radius and is typically smaller. For example, for the parameters shown in Fig.~\ref{fig:S_curve}, when $\eta=0$ we find $|\Delta M^\pm|/M^\pm \lesssim 5\%$.

Further examination of Eq.~\eqref{eq.Mdot_crit_mag} shows that for positive $g$,  a situation may arise in which the critical values $\Mdot^\pm_{\rm magn}$ become arbitrarily small at radii close to $r_0$. This means that the disc temperature remains above the critical value for any accretion rate because of the torque and energy transferred from the rotating central star, and the disc material cannot recombine. Let us define $r_{\rm hot}$ as the maximum radius at which this can occur. Substituting $\dot{M}^+_{\rm non-magn} = 0$ into Eq.~\eqref{eq.Mdot_crit_mag}, we obtain
\begin{equation}
\label{eq.rhot}
r_\mathrm{hot} ~[\mathrm{cm}] = 
    \begin{cases}
    \displaystyle
    2.2\times10^9 \,(\mu_{26}f_{200})^{0.63} M_{1.4}^{-0.2} \left(\frac{g}{\omega^2}\right)^{0.31},&\mathrm{AMXP,~}\\
    \displaystyle
    1.1\times10^{10} \, \left(\frac{\mu_{30}}{P_{4~s}}\right)^{0.63} M_{1.4}^{-0.2} \left(\frac{g}{\omega^2}\right)^{0.31},&\mathrm{XRP, ~} \\
    \displaystyle
    4.1\times10^{10} \,\left(\frac{\mu_{33}}{P_{\rm 10~min}}\right)^{0.63} M_{0.7}^{-0.2} \left(\frac{g}{\omega^2}\right)^{0.31},&\mathrm{IP. ~} \\
    \end{cases}
\end{equation}

\subsection{Radial profiles}\label{sec:rad_str}
\begin{figure*}
\includegraphics[width=\textwidth]{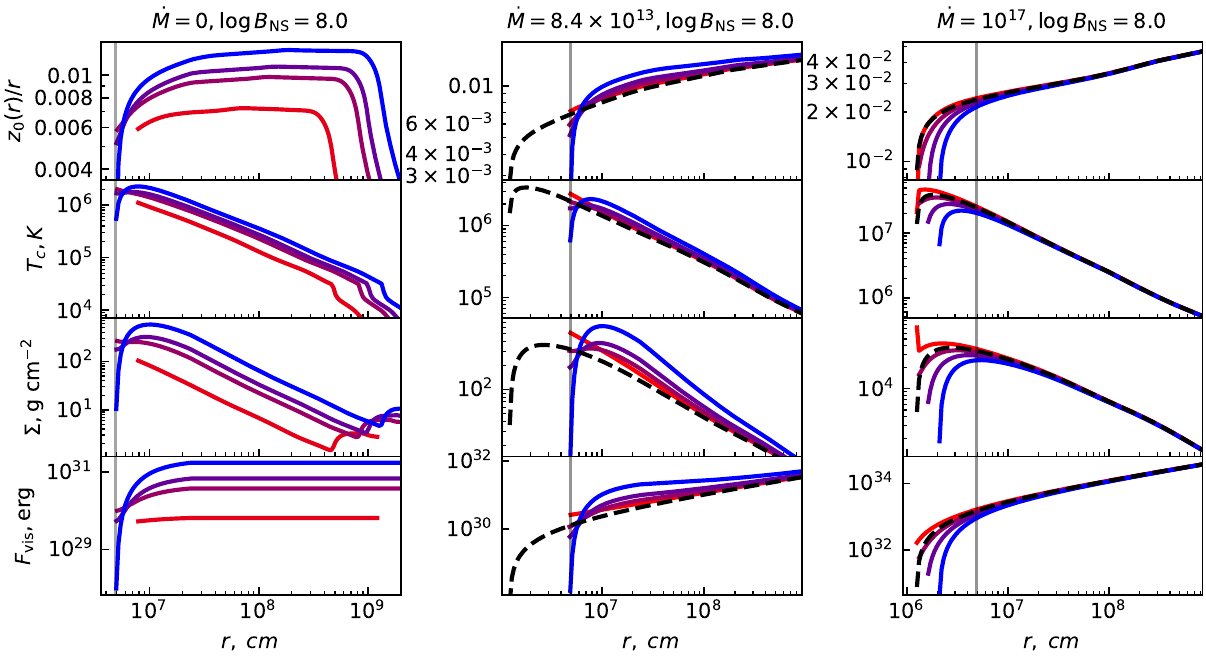}
 \caption{Disc radial structure. The rows correspond to: (1) the relative disc half-thickness $z_0/r$, (2) the central temperature $T_{\rm c}$, (3) the disc surface density $\Sc$, and (4) the viscous torque $F_\mathrm{vis}$. Parameters typical to an AMXP are adopted, with $f=200\mathrm{~Hz}$ and $B_\mathrm{NS}=10^8$~G. The columns correspond to different accretion rates: the first column shows the zero-accretion case, $\Mdot= 0$; the second column corresponds to the accretion rate for which $r_0(\eta=0)=\rc$; and the third column demonstrates the strong accretion case, $\Mdot = 10^{17} \mathrm{g}~\mathrm{s}^{-1}$. Vertical gray lines indicate the corotation radius. The inclination angle is $\chi = 30^{\circ}$ throughout. In the first column, the colours correspond to $\eta=1$ (blue), $0.6$, $0.4$, and $0.1$ (red). In the other columns, the colours correspond to $\eta=1$ (blue), $0.6$, $0.4$, and $0$ (red). Black dashed curves in the second and third columns show the non-magnetic standard $\alpha$-disc solution. When the central temperature reaches values corresponding to hydrogen recombination, $\sim\mathrm{a~few}\times10^4$~K, the parameter profiles become non-monotonic because of the strong temperature dependence of the opacity coefficient; see, for example, the curves in the left column.
 }
 \label{fig:rad_paper_new}
\end{figure*}

By solving for the vertical structure at different distances from the star, we obtain the radial structure of the disc. In Fig.~\ref{fig:rad_paper_new}, we show the radial profiles of the relative half-thickness $z_0/r$, the central temperature $T_{\rm c}$, the surface density $\Sc$, and the viscous torque for a typical AMXP. We explore how the radial structure behaves for various accretion rates and, in particular, calculate the structure of a disc with zero accretion rate, in which the torque at the inner edge of the disc is generated by its interaction with the magnetosphere. For each model, we consider a range of values of the parameter $\eta$.

In the strong-accretion regime, as $\eta$ increases, the disc is not only truncated at a larger radius (see Fig.~\ref{fig:r0}), but also becomes colder, thinner, and less dense than the standard $\alpha$-disc. This is illustrated in the right column of Fig.~\ref{fig:rad_paper_new};
the magnetic torque partly replaces the viscous torque in transporting angular momentum, increasingly so for larger $\eta$. Consequently, viscous heating is reduced ($g(x)<0$ when $\omega \ll 1$, see  Fig.~\ref{fig:g_vs_x}).
Additionally, the value of $\eta$ affects  the transition  to radiation-pressure regime, an effect which, however, we do not consider in the present work.

In the weak-accretion regime, the situation is reversed: as $\eta\rightarrow 1$, the disc starts being truncated closer to $r_0 \sim \rc$ and gains and transports more angular momentum from the rotating magnetosphere. As a result, the disc gets hotter than a standard disc would be, which is shown in the middle column of Fig.~\ref{fig:rad_paper_new}. In the left column of the Figure, we show the zero-accretion-rate case ($\Mdot = 0$), where all angular momentum is transferred by the magnetic torques (see also  the end of the previous subsection).

{Squeezing of the diamagnetic disc near the inner radius (described by the factor $D$, see Eq.~\eqref{coef_D}) leads to the characteristic behaviour of the red curves close to $r_0$, as seen in the right column of Fig.~\ref{fig:rad_paper_new}. However, we should note that
the disc structure close to $r_0$ may be sensitive to parameter $\delta_0$,
an effective disc aspect ratio at the inner edge, prescribed as $\delta_0 = 0.01$ rather than calculated self-consistently. In addition, as already pointed out, our calculation neglects radiation pressure, even though it becomes significant near $r_0$ for $\eta=0$ and $\dot M \gtrsim 10^{16}$~g~s~$^{-1}$.}

\section{Discussion}\label{sec:colcl}
In the present work, we assume that the induced magnetic field is known apriori. This `kinematic approximation' is justified by the fact that the only parameter of the magnetic field obtained in Paper I that affects the induced field and is not known prior to the calculation of the vertical structure is the relative disc height, $\delta(r) = z_0(r)/r$, and that $b(z, r)$ weakly depends on $\delta$ (see Fig.~\ref{fig:magn}).

The `kinematic approach' also implies that $\alpha$, and therefore the magnetic diffusivity, is constant along the vertical coordinate. However, the turbulent parameter $\alpha$ may vary with $z$. In particular, this was shown by \citet{Hub21} through modelling of the Balmer jump in spectra of nova-like variables. On the other hand, \citet{GO13} showed that a height-dependent profile $\alpha(z)$, motivated by numerical simulations \citep{FLM11, ZS18}, does not qualitatively alter the pressure, temperature, and related profiles. A self-consistent treatment of the magnetic-diffusion equation together with the equations of vertical structure was attempted by \citet{Khai22}.

Magnetic buoyancy is another potentially important effect that we have not considered. This effect may significantly influence the induced field inside the disc when the magnetic and gas pressures become comparable, as pointed out, for example, by \citet{Coroniti85}. As the field is generated inside the disc, magnetic flux tubes rise towards the surface, carrying magnetic energy and thereby introducing non-locality into the vertical energy dissipation profile. Several authors have attempted to account for this effect and to determine the vertical structure of the disc \citep{Kool99,Begel15}, but in both works the density profiles were postulated apriori, making direct comparison with our results difficult.

Reconnection of magnetic field lines above the disc can also limit the induced field, since the lines may reconnect in regions where the Keplerian velocity in the disc differs substantially from the magnetospheric velocity \citep{Lovelace95}. Reconnection affects the radius $r_\mathrm{out}$ (see \S\ref{s.conductivedisc} and Paper I) and may be accounted for by manually setting $r_{\rm out} < 2^{2/3}\rc \approx 1.59 \,\rc$, such that the induced field remains smaller than the vertical component everywhere: $b_{\fim}/B_z < 1$.

The vertical structure of an accretion disc with an induced magnetic field is also of interest as a necessary ingredient for studying jet launching and the physics of winds and outflows from the disc. The structure of the disc, accounting for the radial and vertical velocities $v_r, v_z$, radial and azimuthal induced magnetic fields $b_r, b_\fim$, and deviations of the azimuthal velocity from the Keplerian law, $\delta v_\fim$, has been investigated by several authors \citep{OL01, GO12, GO13, LRBK09, BKL12}. Since the primary focus of these works was the vertical profiles of velocity and magnetic field, the energy-transport equation was not considered in most of them, except for \citet{OL01}, and instead a simplified isothermal or adiabatic equation of state was adopted.

In our work, we restrict ourselves to the Keplerian disc approximation (see Appendix~\ref{subsec:omega}), which appears reasonable because all the aforementioned works found $\delta v_\fim / v_{\rm k} \ll1$. We also neglect the radial induced magnetic field, although some of the cited works found that it may be of the same order as $b_\fim$ \citep{OL01, LRBK09, BKL12}.

\subsection{Observational predictions}
\subsubsection{Dependence of the inner radius on the accretion rate}\label{subsec:pds}
The location of the inner disc radius can be studied by examining the power density spectra (PDS) of the aperiodic flux variability of accreting pulsars. These PDS are explained by the `perturbation propagation' model \citep{Lyub97}. The typical PDS, $P_\nu$, of a source containing a WD or an NS consists of two power laws with a knee at a characteristic `break frequency' $\nu_{\rm b}$, which is often associated with the Keplerian frequency at the inner radius \citep{Revn09}:
\begin{equation}\label{nu_break}
    \nu_{\rm b} = \frac{1}{2\pi}\sqrt{GM/r_0^3}.    
\end{equation}

\begin{figure}
\includegraphics[width=\columnwidth]{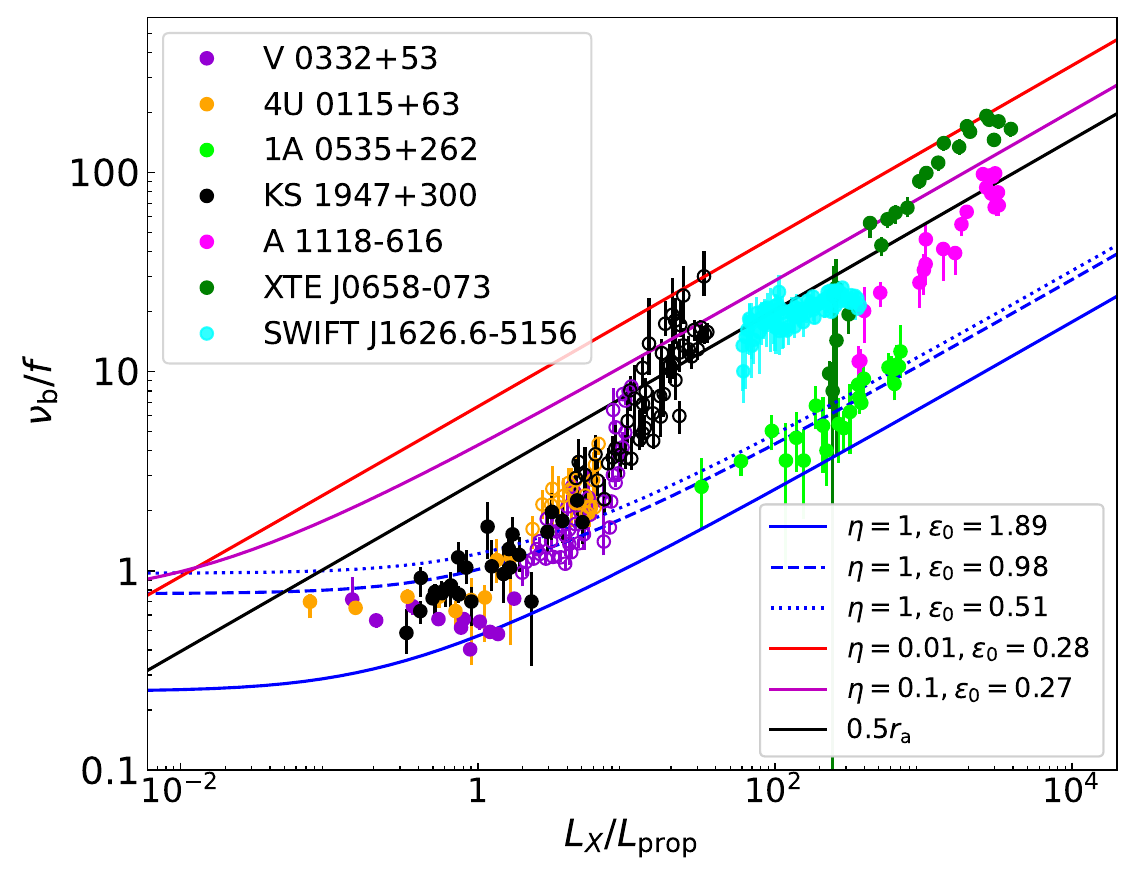}
 \caption{Observed break frequency (normalized by the pulsar spin frequency) as a function of the source X-ray luminosity (in units of the propeller luminosity; see text) for seven accreting Be/X-ray pulsars. The data are taken from Fig.~11 of \citet{Reig13}; filled and hollow markers correspond to subcritical and supercritical accretion, respectively. The flux values from the cited paper were rescaled using updated distances from the HMXBcat catalogue \citep{Neun23}. For KS~1947+300, the magnetic field is not known \citep[see the Introduction of][]{Liu23}, so the value $E_\mathrm{cyc}=50$~keV was adopted. Alongside the data, theoretical predictions are shown for several values of $\eta$ and $\varkappa$, assuming that the break frequency is equal to the Keplerian frequency at the inner edge of the disc. For all models, $\chi=60^\circ$ and $\delta_0=0.1$.
 }\label{fig:V0332}
\end{figure}

In Fig.~\ref{fig:V0332}, we show the dependence of the break frequency, normalized by the spin frequency, on luminosity, using data from \citet{Reig13} for seven pulsars with Be companions, together with the corresponding theoretical predictions. We normalize the observed luminosity by $L_\mathrm{prop}\equiv 0.2 \Mdot_{\rm prop} c^2$, which renders the theoretical curves independent of the pulsar spin frequency and magnetic field. The theoretical predictions are shown for several values of $\varepsilon_0$ and $\eta$, assuming that the break frequency is equal to the Keplerian frequency at the inner radius.

For two sources, V 0332+53 and 4U 0115+63, the observed $\nu_\mathrm{b}$ gradually approaches a fraction of the pulsar frequency $f$, which can be explained by  theoretical torque models with $\eta \sim 1$.

However, it is worth noting that the break frequency does not necessarily coincide with the Keplerian frequency at the inner radius. Indeed, \citet{Monk22} analysed approximately 15 accreting objects of various types, including AMXPs, XRPs, and IPs, and found a noticeable discrepancy between the theoretical and observed dependence $\nu_{\rm b}(\mu)$ for XRPs (see their Fig.~2), where the theoretical relation was calculated under the assumption $r_0 \sim \ra$. Detailed analytical calculations of the PDS for different initial fluctuation spectra \citep{Mush18, Mush19} show that the break frequency $\nu_{\rm b}$ may be smaller than $\nu_{\rm kepl}(r_0)$ by a factor of a few.

To reconcile the observed dependence on accretion rate, which is generally steeper than predicted by models with constant $\eta$, it is necessary for the penetration parameter $\eta$ to decrease  as $\dot M$ increases. On the other hand, \citet{Chash19} predict that radiation pressure leads to a weaker dependence of $r_0$ on the accretion rate. This may in fact occur at the high-luminosity end of some data sets shown in Fig.~\ref{fig:V0332}.

The torque model proposed in this paper (see Appendix~\ref{subsec:W} for the explicit formulas) provides a physically motivated framework that can serve as a unified torque model for studying how the nature of the interaction between the disc and the magnetosphere depends on the parameter $\eta$, which itself may vary with accretion rate or source type.

\subsubsection{Validity of the solution near corotation}
{ For  $\eta\sim1$, and $\Mdot\ll\Mdotpr$, the inner radius approaches $\rc\,\omega_{\rm max}^{2/3}$ (see Eq.~\eqref{crit_w_new}). This asymptote is common to threaded-disc models in which the induced toroidal field is proportional to $(\omegak - \omegas)$ \citep{Wang95, Wang97, KR07}: the inner edge condition \eqref{eq_r0_eta_app} contains the factor $(1-\omega)$ and can be met only by moving $r_0$ towards $\rc$. Prescriptions with a fixed magnetic pitch at $r_0$ \citep{Wang1996,Wang97} lack this factor and give $r_0\propto\ra$ instead (Fig.~\ref{fig:r0comp}). For the same critical fastness, the constant-pitch approximation and the full solution of the torque-balance equation give very different $r_0(\Mdot)$ near $\rc$ \citep[see][]
%[Sect.~2.2.1 and Fig.~1]
{Bozzo09}. Since an expression for the toroidal field is, largely, an ad-hoc assumption,  the precise shape of $r_0(\Mdot)$ should not be over-interpreted.

We review here which assumptions of our model can break when the inner radius of the disc is close to the corotation one. First, the  Keplerian approximation  remains largely valid. The change of the shear factor caused by $\delta_\omega$ (Appendix~\ref{subsec:omega}) becomes comparable to $1-\omega$ only for $\dot m\lesssim\eta^2(f_\chi\,\alpha\,\delta_0^3)^{1/2}\sim10^{-3}$. 

More restrictive is the treatment of the inner boundary and the conditions there. Our model resolves no transition region between the Keplerian and rigidly rotating  flow. As in the model of \citet{Wang87, Wang95}, $\eta$ is constant from $r_0$ outwards and the screening currents are not computed self-consistently, so the conditions at $r_0$,  Eqs.~\eqref{New_cond_W} and \eqref{aux_r0_1}, are effective conditions for the disc outside the unresolved transition zone. They act over the scale $z_0(r_0)=\delta_0 r_0$: this is the radial scale over which the magnetic pressure jump \eqref{delta_Pmagn_r} (and, for $\chi\neq0$, the non-axisymmetric field $b_1$, see Eq.~\eqref{b1_r}) is applied. The conditions remain meaningful only while $\rc-r_0\approx\frac{2}{3}(1-\omega)\,\rc$ exceeds this scale, that is, for $\dot m\gtrsim3\eta^2f_\chi\,\delta_0$ ($\approx0.03$--$0.2$ for $\delta_0=0.01$--$0.05$; cf.~\citealt{Bozzo09}, Sect.~5, for the boundary layer of \citealt{GL79a}). At lower rates, the model indicates only that $r_0\approx\rc$. 

A robust result is  the fastness at which the torque reverses: $\omega_{\rm eq}\approx0.88$--$0.95$ ($0.96$--$0.99$ for $r_{\rm out}=2^{2/3}\rc$) for $\chi=0^\circ$--$45^\circ$, because this is set by the radial profile of $b_\fim B_z$ (Appendix~\ref{subsec:W}). However, the value of the accretion rate at which this happens, $\dot m\approx0.02$--$0.3$ ($0.02$--$0.08$), can  only be said to lie near this limit; it shares the uncertainty of the inner-radius behaviour with decreasing $\dot M$, see above.

Finally, our model assumes a non-zero accretion rate through the inner boundary, $\Mdot>0$, which renders solutions with $\eta\ll1$ and $\Mdot\ll\Mdotpr$ physically questionable. 
For $\omega>1$, the efficiency of matter ejection ($f_\star<1$), its dependence on $\eta$, and the structure of the transition zone require a dedicated treatment~\citep[e.g. by means of MHD simulations, as][]{Romanova+2018}. The classical boundary layer \citep[][who obtained $\eta\approx0.2$]{GL79a, GL79b}, in which the magnetic stresses decelerate the plasma towards corotation and channel it onto the star, has no steady solution for $\omega\gtrsim0.95$ \citep{GL79b} and requires $\omega<1$.
%; for fast rotators its width becomes comparable to $\rc-r_0$ \citep[][Sect.~5]{Bozzo09}.
For $\eta\sim1$, our model resembles a disc-trapped model~\citep{Dangelo-Spruit2010}, in which efficient interaction of the disc matter and the magnetic field ensures that the inner edge of the disc never moves far away from the corotation radius, %$r_0\rightarrow\rc\,\omega_{\rm max}^{2/3}\approx\rc$ as $\Mdot\rightarrow0$ (Eq.~\eqref{crit_w_new}), 
and the limit $\Mdot=0$ is the dead disc with a non-zero magnetic torque at $r_0$.
}

\subsubsection{Transitional millisecond pulsars}\label{subsubsec: transit}
The presented disc model has a close connection to accretion models proposed for transitional millisecond pulsars~\citep{Papitto-deMartino2022}. Observations of X-ray pulsations during low-luminosity states indicate ongoing accretion onto the NS magnetic poles. The inferred classical value of the magnetospheric radius implies that it exceeds the corotation radius, which makes it difficult to explain continued accretion onto the star. It has been suggested that the actual accretion rate is high, allowing the disc to approach $\rc$, while efficient outflows near the centrifugal barrier prevent most of the mass from accreting onto the star~\citep{Papitto-Torres2015}. Alternatively, an inclination-dependent inner disc radius was proposed by \citet{bozzo}.

For the accretion rate approaching zero  \citep[also known as `dead disc' configuration suggested by][]{SS77}, $\Mdot \rightarrow 0$, the magnetic torque acting on the disc is equal to the torque spinning down the star:

\begin{equation}
    \tau_{\rm \star \,\,magn}  = - \frac{\mu^2}{ \rc^3} \, \frac{g}{\omega_{\rm max}^2} \, , \quad \textrm{where} \quad g\approx \frac{\eta^2 \cos^2\chi}{C} \, 
    \label{eq.tau_limit}
\end{equation}
for $\eta$ not too small and $\chi$ not too close to $\pi/2$, where $\omega_{\rm max}\sim 1$ is defined by Eq.~\eqref{crit_w_new}. The dimensionless parameter $C\geq 3$ depends on the radial extent of the disc--magnetic field interaction region (see Appendix~\ref{subsec:W} and Eq.~\eqref{eq.tau_star_eta1}).

The maximum spin-down rate of the central star is
\begin{equation}
    \label{eq.spin_down_rate}
    \dot \nu \approx - \frac{\mu^2}{2\, \pi\, I\, r_0^3}\,\, g = - 8.5 \times 10^{-14} {\rm s} ^{-2}~ \frac{g}{\omega^2}\, \frac{R_{6}^6 B_8^2\,}{P_{\rm ms}^{2} \,I_{45}\, M_{1.4}}  \, ,
\end{equation}
where $R_6=R_\star/10^6$~cm, $P_{\rm ms} = P_\star/10^{-3}$\,s, $I_{45}=I/10^{45}~$g cm$^2$.  The general expression for the torque is given by Eq.~\eqref{eq.n_general}.

The spin-down \eqref{eq.spin_down_rate}  can be higher than the spin-down rate of the radio-pulsar mechanism, $\dot \nu_{\rm PSR}  \sim - 8 \pi^2 \mu^2 \sin^2 \chi /(3\,  I c^3 P_\star^3) $:
\begin{equation}
 \frac{\dot \nu}{   \dot \nu_{\rm PSR}}  \sim 10 \, P_{\rm ms} \, \frac {3 \, \eta^2}{C \tan^2{\chi}}\, ,
\end{equation}
which is characteristic of the  sub-luminous state of transitional AMXPs~\citep{Papitto-deMartino2022}. 

The energy released close to the disc inner edge $L \sim \omegak(r_0) \,\mu^2\,g\,/r_0^3$ is  
\begin{equation}
    \label{eq.L_dead}
     L\sim  \frac{\mu^2\sqrt{GM}}{r_c^{9/2}}\,\frac{g}{\omega^3}\, = 3 \times 10^{36} {\rm erg\, s}^{-1}~ \frac{g}{\omega^3}\,\frac{R_{6}^6 B_8^2\, }{P_{\rm ms}^{3}\, M_{1.4}} \,.
\end{equation}
It depends strongly on $\eta$, since for $\eta\rightarrow 1$ the disc can remain just outside the corotation radius, ensuring that $\omega\sim 1$. 

The thermal spectrum of such a disc differs from that of a strongly accreting disc. A known feature of `dead discs', pointed out previously, is the weak dependence of the viscous torque on radius, which leads to an intermediate thermal spectral index of $5/7$ \citep{SS77}, in contrast to the index $1/3$ for a standard disc \citep{Lyn69, SS73}. While the maximum effective temperature of such disc,  estimated from $\sigma T_{\rm eff}^4 \pi \rc^2 = L$,  is about $0.7~g^{1/4}\, P_{\rm ms}^{-13/12} B_8^{1/2}$~keV,  possible high-energy processes at the magnetosphere boundary can produce power-law spectra. 

\section{Summary}\label{sec:summary}
The aim of the present work was to calculate the structure of a stationary accretion disc around a magnetized star. The equations of motion for matter in the disc were modified appropriately to account for the magnetic field penetrating the disc. Since the details of the interaction between the disc and the magnetic field remain uncertain, we introduced the parameter $\eta$, with $0 \leq \eta \leq 1$, representing the fraction of the vacuum dipole field of the star that remains inside the disc.

The introduction of the parameter $\eta$ unifies two approaches to the disc structure: the `diamagnetic disc' ($\eta=0$) and the disc with effective turbulent diffusivity, the `conductive disc' ($\eta > 0$). The `diamagnetic disc' approach leads to the inner radius of the viscous accretion disc being of the order of the Alfven radius, $r_0 \sim \ra$ (Eq.~\eqref{r0_eta0}). The `conductive disc' approach results in $r_0$ confined within a few corotation radii: $r_0 \lesssim \textrm{a few~}\rc$ (Eq.~\eqref{eq.r0_new}), a conclusion similar to those previously reached by \citet{KR07}, \citet{Wang95}, \citet{Wang97}, and \citet{bozzo}. In our model, the maximum possible inner radius, or equivalently the maximum fastness parameter, is defined by Eq.~\eqref{crit_w_new} and depends on the penetration factor $\eta$ and the magnetic inclination angle $\chi$.

The magnetic field was shown to significantly influence the viscous heat release rate by modifying the viscous stresses (Eq.~\eqref{eq.wrf_wang}). For `conductive discs' with $\eta \sim 1$ and weak accretion, when the inner radius satisfies $r_0 \sim \rc \ll \ra$, the magnetic stresses distributed over the disc radius dominate the viscous heating rate, and the disc tends to be thicker and hotter than a standard disc (see, e.g. Fig.~\ref{fig:rad_paper_new}). In case of strong accretion in a `conductive disc' model, or if the `diamagnetic disc' model  is adopted ($\eta\sim 0$ and any $\Mdot$), the disc is either indistinguishable from the standard disc or colder and thinner.

The structure can be calculated using the open-source code \alphadisc. We found vertical structure of a disc embedded in an external magnetic field and found that the vertical profiles of pressure, temperature, and flux, when normalized by their characteristic values, are approximately universal (see Fig.~\ref{fig:vert_struct}). We were also able to calculate the disc structure at radii much smaller than the Alfven radius $\ra$, where the magnetic field substantially modifies the viscous stress tensor. Another potentially important finding is that the value of $\eta$ affects the extent of the central radiation-pressure-dominated zone.

For magnetically threaded discs, the interaction between the disc matter and the magnetosphere modifies the conditions for thermal and viscous instabilities compared to the non-magnetic case. The equatorial and effective disc temperatures, as well as the disc surface density at the turning points of the S-curves, remain unchanged, whereas the critical accretion rate may be significantly altered (see Fig.~\ref{fig:S_curve} and Eq.~\eqref{eq.Mdot_crit_mag}). 

In contrast to diamagnetic disc models with $r_0\sim \ra$, for magnetically threaded discs there may exist a maximum radius $r_\mathrm{hot}$ (see Eq.~\eqref{eq.rhot}) such that the disc always remains in the hot state for $r<r_\mathrm{hot}$. Our model for threaded discs also yield specific predictions on the form of the PDS (section \ref{subsec:pds}) and means to provide enhanced spin-down rate in the regime of weak accretion (section \ref{subsubsec: transit}).

\begin{acknowledgements}
This work was supported by Deutsches Zentrum f\"ur Luft- und Raumfahrt e.V. (DLR) grant 50OR2409 (AK) and by  the Deutsche
Forschungsgemeinschaft (DFG, German Research Foundation), project
number \mbox{570950648} (GL). 
\end{acknowledgements}

%%%%%%%%%%%%%%%%%%%%%%%%%%%%%%%%%%%%%%%%%%%%%%%%%%%%%%%%%%%%%%
% WARNING
% Please note that we have included the references below in
% order to compile the document, but we ask you to:
%
% - use BibTeX with the regular commands:
\bibliographystyle{aa} % style aa.bst
\bibliography{aa} % your references Yourfile.bib
% - join the .bib files when you upload your source files
%%%%%%%%%%%%%%%%%%%%%%%%%%%%%%%%%%%%%%%%%%%%%%%%%%%%%%%%%%%%%%

%%%%%%%%%%%%%%%%%%%%%%%%%%%%%%%%%%%%%%%%%%%%%%%%%%%%%%%%%%%%%%%
% Appendices must be placed after   \end{thebibliography}
% They will be placed automatically on a new page.
%%%%%%%%%%%%%%%%%%%%%%%%%%%%%%%%%%%%%%%%%%%%%%%%%%%%%%%%%%%%%%%
\begin{appendix}

\section{$\alpha$-disc embedded in external magnetic field}\label{App:general}
We write the external dipole magnetic field 
in the form
\begin{multline}
        \vb{B}^{\rm dip} =\\
        \frac{\mu}{(r^2 + z^2)^{5/2}}  [(2r^2 \sin{\chi} \cos{\varphi'} + 3rz \cos{\chi} - z^2 \sin{\chi} \cos{\varphi'})\, \vb{e}_r +\\
        + (\sin{\chi}\sin{\varphi'} (r^2 + z^2))\,\vb{e}_{\varphi'} + \\
        + (-r^2\cos{\chi} + 3rz \sin{\chi}\cos{\varphi'} + 2z^2 \cos{\chi})\,\vb{e}_z]\, .
        \label{dipole_field_full}
\end{multline}
For the case of small heights above the equatorial plane,  $z \ll r$, it reduces to Eq.~\eqref{dipole_field}.

\subsection{Magnetic field inside and outside the disc}\label{App:MF}
In \S\ref{s.intermediate}, we consider the situation in which the time-independent component of the external dipole magnetic field partly penetrates the disc. The dominant terms in $\vb{B}_r^{\rm dip}$ and $\vb{B}_\fim^{\rm dip}$ are time-dependent (i.e. depend on $\varphi'$) and are therefore screened out from the disc, except in the vicinity of the inner edge of the disc (see below). Thus, for $r > r_0$, just above and below the disc surface, with $\varepsilon\ll 1$, we have
\begin{equation}
    \begin{cases}
    z = z_0\,(1 + \varepsilon):\\
    \displaystyle
        B_r = B_r^{\rm dip} + \frac{4}{\pi D}  (1 - \eta)B_z^{\rm dip}\\
    \displaystyle
        B_{\fim} = B_{\fim}^{\rm dip} +\eta b_{\fim} \\
    \displaystyle
        B_z = \eta B_z^{\rm dip}\, ,
    \end{cases}\qquad
       \begin{cases}
       z = z_0\,(1 - \varepsilon):\\
    \displaystyle
        B_r = 0  \\
    \displaystyle
        B_{\fim} = \eta b_{\fim}\\
    \displaystyle
        B_z = \eta B_z^{\rm dip}\, ,
    \end{cases}
    \label{eq.B_eta_full}
\end{equation}
where $b_{\fim}$ is the induced field corresponding to the case in which the dipole field penetrates the disc without screening; cf.~Eq.~\eqref{bphi_small}, and $D$ is defined by Eq.~\eqref{coef_D}. The rotation of the disc matter induces the toroidal component $\eta b_\varphi$ both inside and outside the disc. The $z$-component of the magnetic field is continuous across the interface.

At the inner radius, we also account for the radial time-dependent component of the dipole field, which partially penetrates the disc and extends only a short distance into it: 
\begin{equation}
       \begin{cases}
       r = r_0\,(1 - \varepsilon):\\
    \displaystyle
        B_r = \eta B_r^{\rm dip} \\
    \displaystyle
        B_{\fim} = B_{\fim}^{\rm dip} +\eta b_{\fim} \\
    \displaystyle
        B_z = \eta B_z^{\rm dip} + \frac{4}{\pi  D(r_0)} (1-\eta)\, B_z^{\rm dip} ,
    \end{cases} \qquad
    \begin{cases}
    r = r_0\,(1 + \varepsilon):\\
    \displaystyle
        B_r = \eta B_r^{\rm dip}\\
    \displaystyle
        B_{\fim} = \eta b_{\fim} \\
    \displaystyle
        B_z = \eta B_z^{\rm dip}\, ,
    \end{cases}
    \label{eq.B_eta_inner}
\end{equation}

It is now straightforward to calculate the discontinuities in the azimuthally averaged magnetic pressure, $\langle \vb{B}^2/8\pi \rangle$, at the disc surface and inner edge, yielding Eqs.~\eqref{delta_Pmagn}~and~\eqref{delta_Pmagn_r}.

 Let us estimate the characteristic 
penetration depth of the non-axisymmetric components inside the disc. This depth may be estimated as the skin depth: 
\begin{equation}\label{skin}
 d_{\rm skin} \approx \left(\frac{c^2}{{2\pi} \sigma |\omegas - \omegak|}\right)^{1/2} = \left(\frac{{ 2} \eta_{\rm T}}{\Delta \Omega}\right)^{1/2} \sim z_0 \left(\frac{\omegak}{\Delta \Omega}\right)^{1/2}\, ,
\end{equation}
where $\sigma$ is the effective conductivity of the plasma, $\sigma = c^2/4\pi \eta_{\rm T}$. Here $\Delta \Omega = |\omegak - \omegas|$, and $\eta_{\rm T}$ is the coefficient of turbulent magnetic diffusivity, assumed in the last equality to be of the order of the kinematic viscosity coefficient, $\alpha \, z_0^2 \, \omegak$, while dropping the factor $\sqrt{{2}\alpha}$. At the inner radius, the factor $(\omegak / \Delta \Omega)^{1/2} = |1 - \omega|^{-1/2}$. We see that the non-axisymmetric component may penetrate to depths compatible to the scale height inside the disc: for example, $d_{\rm skin}/z_0 \approx 1.9$ at $r_0 = 0.8\rc$.  {The numerical prefactor of this estimate is, however, uncertain. 
We therefore adopt $d_{\rm skin} \approx z_0$ as an effective value throughout. Using instead $d_{\rm skin} = z_0\,|1-\omega|^{-1/2}$ changes $r_0$ by $\lesssim 5\%$ for $\chi \leq 45^\circ$ and the equilibrium fastness (see Appendix~\ref{subsec:W}) by $\lesssim 0.05$.} 

We do not take into account the possible twisting of the non-axisymmetric components (such as $B_r^{\rm dip}$) at the upper and lower disc surfaces, since a stationary vertical component of the dipole field is present there and likely dominates. \citet{Lai99} argued that the field induced at the surface by shearing of $B_r$ is smaller than the field induced by shearing of $B_z$: from the magnetic-diffusion equation it follows that $b_\varphi \propto B_z \pd{\Omega}{z} + B_r \pd{\Omega}{r}$. Hence, $b_\varphi^{\rm from~ B_r} / b_\varphi^{\rm from~ B_z} \sim \delta$.

\subsection{Equations for viscous stress tensor and inner radius}\label{sec:ang_mom}

Using Eq.~\eqref{dipole_field_full}, we average the second equation of system \eqref{main} over $\fim \in [0, 2\pi]$ and integrate it over $z \in [-z_0, z_0]$:
 \begin{multline}\label{eq_phi_main}
- \dot{M} \frac{\dif (\Omega r^2)}{\dif r} = 2\pi \frac{\dif}{\dif r}(r^2 \Wrf) + \frac{1}{2}r^2 \langle B_z B_{\fim}\rangle  \bigg|^{z=z_0}_{z=-z_0} + \\ +\frac{1}{2}\int_{-z_0}^{z_0} \frac{\dif}{\dif r} \langle r^2 B_{\fim} B_r \rangle \text{d}z\, .    
\end{multline}
Introducing the the viscous torque $F_{\rm vis} = 2\pi r^2 W_{r\fim}$, Eq.~\eqref{eq_phi_main} can be written as 
\begin{equation}\label{eq_phi_reduced}
    - \dot{M} \frac{\dif (\Omega r^2)}{\dif r} = \frac{\dif F_{\rm vis}}{\dif r} + 
     \frac{\dif F_{\rm magn}}{\dif r}\, ,
\end{equation}
where $\dif F_\mathrm{magn}/\dif r$ is given by Eq.~\eqref{Fmagn}.
Since $\dif F_{\rm magn}/\dif r = 0$ when $\eta = 0$, the only deviation of the viscous torque from the standard form in the case of a diamagnetic disc arises from the inner boundary condition \eqref{New_cond_W}.

For the magnetic field, the corresponding components from Eqs.\eqref{eq.B_eta_full} (the right column) and \eqref{bphi_small}  can be substituted. In particular, for radii $r\neq r_0$, we write
$\langle B_z B_\fim \rangle = \eta^2 \langle B_z^{\rm dip} (b + b_1 \cos{\fim}') \rangle = \frac{\eta^2 \mu}{r^3}(-b\cos{\chi} + \frac{3z}{2r} b_1 \sin{\chi})$.
We use the approximation $\delta(r) = z_0(r)/r \approx \text{const}$ and the dipole magnetic field given by Eq.~\eqref{dipole_field}. We also use the relations $b(-z) = -b(z)$ and $b_1(z) = \textrm{const}$; see Eqs.~\eqref{eq.brz}~and~\eqref{b1_r}.

The angular velocity is assumed to be Keplerian everywhere. In transforming Eq.~\eqref{eq_phi_main}, we neglect all terms $\propto z\times \mathcal{O}(z_0/r) = o(z_0/r)$, as elsewhere in this work. In any case, if the disc were not geometrically thin, it would be necessary to incorporate non-Keplerian rotation, radial pressure gradients, and energy advection.

We also interchange the order of integration and differentiation: $\int_{-z_0}^{z_0} \frac{\dif}{\dif r} \rightarrow \frac{\dif}{\dif r} \int_{-z_0}^{z_0}$. This is not entirely self-consistent, since the thickness depends on radius, $z_0 = z_0(r)$. However, this approximation has the advantage that the last term becomes the total derivative $\frac{\dif}{\dif r}(b_1 r^{3/2} (z_0/r))$ instead of $z_0 \frac{\dif}{\dif r}(r^{3/2 }b_1/r)$. This modification proves useful for obtaining simplified expressions for $\Wrf$. \citet{Wang97} employed the formally correct expressions, but neglected the term $\frac{z_0}{r} \frac{3\mu \eta^2 \sin{\chi}}{2r}b_1$ on the right-hand side of Eq.~\eqref{Fmagn}. At the end of this Appendix, we briefly discuss the differences between our final expressions and those of Wang.

Since we adopt the $\alpha$-parametrization for the viscous stress tensor in this work, $w_{r\fim}(r_0)= -\alpha p$, it appears reasonable to set $\Wrf(r_0) = -2\,z_0(r_0)\, \alpha\, \Delta_r P_{\rm m}$ \citep{LyndBellPringle74, AlyKuj90, Chash17, Chash19}, where the magnetic-pressure discontinuity $\Delta_r P_{\rm m}$ at the interface is given by Eq.~\eqref{delta_Pmagn_r}. Then $\Wrf(r_0)$ may be written as in Eq.~\eqref{New_cond_W}:
\begin{equation}\label{New_cond_W_app}
   \Wrf(r_0)  \approx -\frac{\mu^2}{r_0^5}\,\frac{2\alpha}{\pi^3}f_P(\eta, \chi, \delta_0)
   = -\frac{\mu^2}{r_0^5}\,\frac{\varkappa'}{\pi}f_P
   \, ,
\end{equation}
where $\varkappa' \equiv 2\alpha/\pi^2$, although we emphasize the approximate nature of this relation. 

The half-thickness $z_0(r_0)$ at the inner edge cannot be calculated within our framework. At the radii where the magnetospheric boundary is usually located, $r_0 \sim 10^7$ cm for AMXPs and $r_0 \sim 10^{10}$ cm for IPs, the standard disc model (with $r_0 = \RISCO$) gives values of $\delta \approx 10^{-2}$ (for AMXPs) and $\delta \approx 2 \times 10^{-2}$ (for IPs) at $\log{\Mdot} \approx 17$. \citet{Chash17} attempted to calculate self-consistently the relative semi-thickness $\delta_0$ at the inner edge of the disc around a magnetized NS and obtained $\delta_0 \sim 10^{-2}-10^{-1}$ for sub-Eddington accretion.

If the inner radius is known, the viscous torque  $F_\mathrm{vis} \equiv 2\pi r^2 \Wrf (r)$, can be calculated from Eq.~\eqref{eq_phi_reduced}, together with Eq.~\eqref{New_cond_W}:
\begin{equation}
\label{Wrphi_new_app}
    F_{\rm vis} (r) =  F_{\rm vis} (r_0) - \dot{M} \Omega_k r^2\left(1 - \sqrt{\frac{r_0}{r}}\right) +
\int_{r_0}^{r}  \dif F_{\rm magn}
    \, .
\end{equation}
Keep in mind that $F_{\rm vis}<0$ and $\dot M>0$.  Equation~\eqref{Wrphi_new} follows directly and a particular analytic form is given by  Eq.~\eqref{eq.wrf_wang}.

We propose the following approach for defining the inner radius of a disc that partially screens the persistent external magnetic field. Let us consider the angular-momentum transfer equation \eqref{eq_phi_reduced} at the inner radius and relate its terms to the fractional contributions to the total mass flux through the inner boundary of the disc:
\begin{equation}\label{aux_r0_1}
    \Mdot =  \Bigg[-\frac{\dif F_{\rm vis}}{\dif h}
    - \frac{\dif F_{\rm magn}}{\dif h}\Bigg]_{r=r_0}\, ,
\end{equation}
where $h = \sqrt{GMr}$ is the specific angular momentum. 
Let us substitute the first term by its approximate value:
\begin{equation}\label{}
   \frac{\dif F_{\rm vis}}{\dif h}\Big|_{r_0} \sim \frac{ 2\pi r_0^2 \Wrf(r_0)}{h_0} = - \frac{4 \alpha \mu^2 f_P}{\pi^2 h_0 \, r_0^3}\, .
\label{eq.dFvis_dh_r0} 
\end{equation}
Here, in the second equality, we used the previously introduced relation~\eqref{New_cond_W_app}.

Further,  from Eq.~\eqref{Fmagn} we obtain
$$
 \frac{\dif F_{\rm magn}}{\dif h}  = - \frac{2\eta^2 \mu}{h}\, (b^\mathrm{u} \cos \chi - \delta_0 \frac{{\rm d} (r^{3/2} b_1)}{r^{1/2} {\rm d} r} \sin \chi)  \, .  
  $$  
Substituting Eq.~\eqref{b1_r} with $d_{\rm skin} = \delta_0 r_0$, see Eq.~\eqref{skin}, we get
$$\frac{1}{\sqrt{r}} \frac{\dif}{\dif r} \left(r^{3/2} b_1(r)\right)\Bigg|_{r=r_0}  = b_1(r_0) (3/2 - r_0/d_{\rm skin})\, .$$
Thus, together with Eq.~\eqref{eq.bur},
\begin{equation}
   \frac{\dif F_{\rm magn}}{\dif h}\Big|_{r_0}  = - \frac{2\eta^2 \mu^2}{h_0 \, r_0^3}\, (1-\omega)\, {f_\chi}\, .
  \label{eq.dFmag_dh_r0} 
\end{equation}
{The inclination factor $f_\chi = 1 + (1-3\delta_0)\sin^2\chi$} can be negative only if $\delta_0 > 1/3$, corresponding to rather thick discs for which our model is not applicable.

Combining equations \eqref{aux_r0_1}--\eqref{eq.dFmag_dh_r0}, we find 
\begin{equation}\label{eq_r0_eta_app}
    \frac{1}{2} \left(\frac{r_0}{\ra}\right)^{7/2} = \varkappa  f_P + \eta^2 (1-\omega)\, {f_\chi}.
\end{equation}

Here, $f_P$ is given by Eq.~\eqref{fp}, and we have substituted $\varkappa = {2\alpha}/{\pi^2}$, noting that this parameter formally has the same value as $\varkappa'$ in Eq.~\eqref{New_cond_W_app}. However, since both arise from approximate arguments, they need not be identical. The parameter $\varkappa$ affects the inner radius, whereas $\varkappa'$ determines the value of the viscous stress tensor at the inner radius. The last relation may be conveniently rewritten by introducing the parameter $\varepsilon_0^{7/2} = 2\varkappa f_P$; see Eq.~\eqref{eq.r0_new}.

We note that the term 
$$(1-\omega)\,{f_\chi}$$
had the form 
$$(1-\omega)\cos^2{\chi} + (8-5\,\omega)\,\delta_0\sin^2{\chi}$$
in \citet{Wang97, bozzo}. The difference is the price paid for the slightly inconsistent but convenient form of Eq.~\eqref{Fmagn}, while retaining all the terms correctly in the equation, see the discussion following Eq.~\eqref{eq_phi_reduced}.  

\subsection{Angular velocity of gas}\label{subsec:omega}
The angular velocity of the gas in the disc may be obtained from the first equation in \eqref{main}. For standard discs, it is customary to neglect $(\vb{v}\nabla)v_r$ and the radial pressure gradient, $\frac{1}{\rho} \pd{P}{r}$, 
both assumptions valid as long as $\delta \ll 1$. This yields the Keplerian angular velocity, $\Omega = \omegak = (GM/r^3)^{1/2}$. Let us briefly examine the effect of the magnetic field on $\Omega$. As a measure of the deviation from Keplerian rotation, we define
\begin{equation}
    \delta_\omega \equiv \frac{\Omega(r_0)^2}{\omegak(r_0)^2} - 1.
\end{equation}

Using the induced magnetic field in the form $b_{\fim} = (z/z_0) ~\mu /r^3$ (see Section \ref{s.conductivedisc}), we obtain an order-of-magnitude estimate:
\begin{equation}
    \delta_{\omega} \sim -\alpha \delta^3 \left(\frac{\ra }{r_0}\right)^{7/2} \approx -m_c(\eta=1, r=r_0) \delta^2, 
\end{equation}
where $m_c\equiv \eta^2 \alpha \delta (\ra/r)^{7/2}$.

This estimate is obtained assuming $\rho \sim \rho_{\rm c}$, where the density in the central plane of the disc, $\rho_{\rm c}$, follows from $\dot{M} = 2\pi r \, 2z_0 \, \rho_{\rm c} v_r$ with $v_r = \alpha \delta^2 v_{\fim}^{\rm kepl}$. We also substitute the disc semi-thickness, $z \sim z_0$.

For a diamagnetic disc, $r_0 \sim \ra$, yielding $|\delta_{\omega}| \sim \alpha \delta^3 \ll 1$. For a magnetically threaded disc, $r_0$ may be of the order of $\rc$, and the deviation from Keplerian rotation increases with decreasing accretion rate: $|\delta_{\omega}| \sim \alpha \delta^3 /\dot m$.

The factor $m_c$ also governs the deviation of the hydrostatic equilibrium from the non-magnetic case, see Sect.~\ref{sec:vert_str}. Similarly, this deviation may become significant for the magnetically threaded disc at low accretion rates:
\begin{equation}\label{eq.m_magn}
    m_c(\eta=1, r=r_c)  \approx 
    \begin{cases}
    \displaystyle
3\times 10^{-5} \,
\mu_{26}^2 \Mdot_{17}^{-1} M_{1.4}^{-5/3}f_{200}^{7/3},&\textrm{AMXP},\\
\displaystyle    
5\times 10^{-4} \,
\mu_{30}^2 \Mdot_{17}^{-1} M_{1.4}^{-5/3}P_{4 ~\rm s}^{-7/3},&\textrm{XRP},\\
\displaystyle    
1 \times 10^{-2} \, \mu_{33}^2\Mdot_{17}^{-1} M_{0.7}^{-5/3}P_{10 \rm ~min}^{-7/3},&\textrm{IP}.
    \end{cases}
\end{equation}

In the present paper, we adopt the Keplerian approximation for the angular velocity throughout. Accordingly, the condition $|\delta_{\omega}|<1$ imposes the following requirement on the magnetic field strength in our model:
\begin{equation}
    \ra / \rc < (\alpha \delta^3)^{-2/7} \approx 25\, \alpha_{0.1}^{-2/7} \delta_{0.05}^{-6/7},
\end{equation}
or equivalently,
\begin{equation}
    \Mdot_{16} \gtrsim 1.5\,\times10^{-3}\,\mu_{30}^2\,P_1^{-7/3}\,M_{1.4}^{-5/3}\,\alpha_{0.1}\,\delta_{0.05}^3,
\end{equation}
which translates to $\Mdot \gtrsim 4\times10^{13}\textrm{~g~s}^{-1}$ for AMXPs and $\Mdot \gtrsim 5\times10^{12}\textrm{~g~s}^{-1}$ for IPs. 

\section{Analytic expressions for viscous torque and spin-up/down torque}\label{subsec:W}

Let us consider how the viscous torque is modified in the disc. Rewriting Eq.~\eqref{eq.wrf_wang} for the viscous stress tensor $\Wrf$ in terms of the viscous torque, $F_\mathrm{vis} \equiv 2\pi r^2 \Wrf(r)$, and moving the terms so that the r.h.s. contains an invariant function for a Keplerian disc, we obtain:
\begin{figure}
    \centering
    \includegraphics[width=1\linewidth]{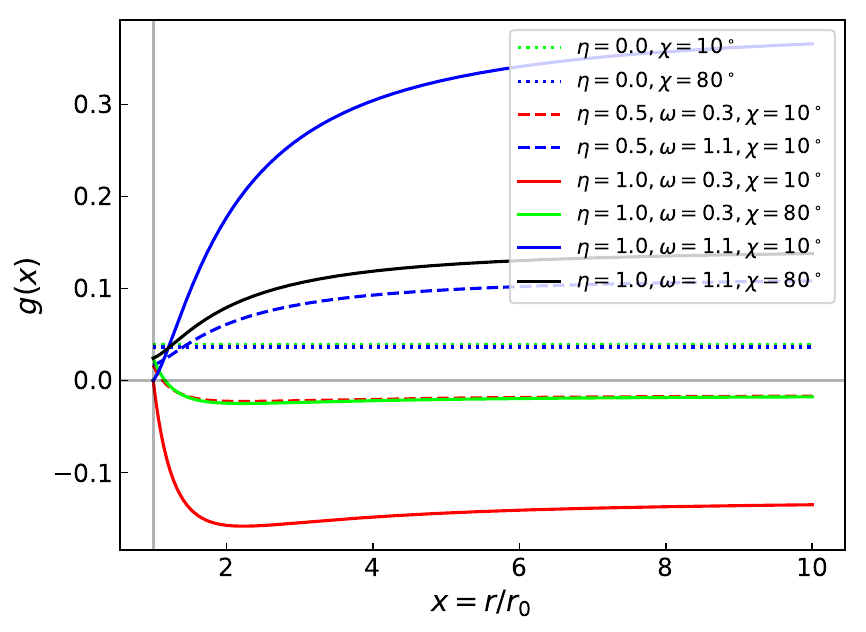}
    \caption{Function $g(x)$ for different values of the fastness parameter $\omega=0.3, 1.1$, penetration parameter $\eta=0, 0.5, 1$, and inclination angle $\chi=10^\circ, 80^\circ$. For all lines, $\alpha=0.1$ and $\delta_0=0.1$. The curves for $\chi=80^\circ$ show the quantity $10\times g(x)$. All dotted lines correspond to $\eta=0$, dashed lines to $\eta=0.5$, and solid lines to $\eta=1$.}
    \label{fig:g_vs_x}
\end{figure}

\begin{equation}\label{F_vis}
     F_\mathrm{vis} + \frac{\mu^2}{r_0^3}\,g(r/r_0) =  - \dot M \sqrt {GMr}\, \left(1-\sqrt{r_0/r}\right) \, ,
\end{equation}
where 
\begin{equation}\label{g_wang}
 g(x) = \frac{\eta^2}{3} \left(2\omega f_{3/2}(x)-f_3(x)\right)\cos^2{\chi} + 2\varkappa' f_P(\eta, \chi, \delta_0)
\end{equation}
with
\begin{equation}\label{eq.fn}
    f_n(x)\equiv 1 - x^{-n}\, .
\end{equation}
Here $\varkappa' = 2\alpha/\pi^2$, see Eq.~\eqref{New_cond_W_app},  $\delta_0$ is the aspect ratio of the disc at the inner edge, and it is assumed that $x < r_{\rm out}/r_0$. The function $f_P$ is given by Eq.~\eqref{fp}. For $x > r_{\rm out}/r_0$, the integration in Eq.~\eqref{Wrphi_new} reduces to $\int_{r_0}^{r_{\rm out}} b^\mathrm{u}(r)/r \,\dif r$, and $g(x) \equiv g(r_{\rm out}/r_0)$.

For $r\gg r_0$, the function $g(x)$ becomes constant.
This reflects the fact that the interaction between the disc and the magnetic field rapidly weakens with increasing radius, so that no additional torque is applied to the disc at large radii.  The behaviour of the function $g(x)$ for several parameter sets is shown in Fig.~\ref{fig:g_vs_x}. For any value of $\eta$, the magnetic field contribution to the viscous stress is constant at large distances.

\begin{table}
\caption{Selected extremal values of the dimensionless torque $n_\omega$. 
}\label{tab.nw}
\centering
\renewcommand{\arraystretch}{1.25}
\begin{tabular}{c |c |c| c c}
\hline
$r_{\rm out}/\rc$ & \multicolumn{2}{c}{}  &  {$2^{2/3}$}&
$\infty$ \\
\hline 
\multirow{5}{*}{$n_\omega$} &
\multirow{3}{*}{$\eta \rightarrow 1$} & 
\makecell{$\chi=0,$\\$\omega\rightarrow 0$} & 
\multicolumn{2}{c}{$f_\star+\dfrac{1}{6}$} \\
\cline{3-5}
 &
&\makecell{$\chi=0,$\\$\omega\rightarrow 1$} & 
$  -\,\dfrac{1}{12\dot m}$ &
$  -\,\dfrac{1}{3\dot m}$ \\
\cline{2-5}
& $\eta \rightarrow 0$ &
\multicolumn{3}{c}{$f_\star -  \dfrac{\varkappa'}{\varkappa}$} \\
\hline
\end{tabular}
\end{table}

In the context of spin-up and spin-down of the central star, our results resemble the ones obtained earlier by \citet{KR07,Wang87,Wang1996}, among others. The total magnetic torque acting on the accretion disc is given by the second term in Eq.~\eqref{F_vis}, evaluated in the limit $r\rightarrow \infty$. The torque acting on the star is the sum of the negative magnetic torque acting on the disc and the material torque \citep[neglecting the stellar rotation; see][]{KR07}:
\begin{equation}\label{eq.tau_general}
    \tau_\star = f_\star \Mdot \sqrt{GM r_0} - \frac{\mu^2}{r_0^3}\,g(r_{\rm out}/r_0) .
\end{equation}

Here, we take into account that only a fraction $f_\star$ of the angular momentum (or accretion rate) reaches the central star. In particular, in high-fastness regimes, outflows from the inner disc radius are possible, leading to $f_\star < 1$.  If $r_\mathrm{out} \rightarrow \infty$, then $f_{3/2} = f_{3} = 1$ at $r_\mathrm{out}$. If instead $r_\mathrm{out} = 2^{2/3} \, \rc$, such that the induced field never exceeds the vertical component, $b_\fim\leq B_z$, then the values of $f_{3/2}$ and $f_3$ at $r_{\rm out}$ become $1-\omega/2$ and $1-\omega^2/4$, respectively.
   
It is customary to define the dimensionless torque $n_\omega$ as follows:
\begin{equation}\label{eq.n_general}
    \tau_\star  \equiv \Mdot \sqrt{GM r_0} \, n_\omega .
\end{equation}

Various forms of $n_\omega$ can be found in the literature; our result is more general in the sense that it depends explicitly on the magnetic inclination angle $\chi$ and the penetration parameter $\eta$. The dimensionless torque $n_\omega$ may be written as follows:
\begin{equation}\label{full_nw}
     n_\omega =f_\star- \frac{g(r_{\rm out}/r_0)}{(r_0/\ra)^{7/2}}  \, .
 \end{equation}
 Substituting $r_0$ from Eq.~\eqref{eq.r0_new},
 \begin{equation}\label{full_nw_subst}
     n_\omega =f_\star- \frac{g(r_{\rm out}/r_0)/2}{\varkappa f_p +  \eta^2 \,(1-\omega)\, {f_\chi} }\, ,
 \end{equation}
where, as before, ${ f_\chi = 1 + (1-3\delta_0)\sin^2\chi\, ,}$
$$
f_P = \left((1-\eta)^2 + \eta~(1-\eta) ~\pi \sqrt{\frac{\delta_0}{2}}\right)\cos^2{\chi}  + \frac{\pi^2\delta_0}{16} \sin^2{\chi} \, ,
$$
and the definitions of $\varkappa$ and $\varkappa'$ are given after Eqs.~\eqref{New_cond_W_app}~and~\eqref{eq_r0_eta_app}.

For a disc with a high degree of magnetic field penetration, $\eta\rightarrow 1$, we recover the expected spin-up and spin-down torques depending on the accretion rate. In the strong-accretion regime with low fastness $\omega$, one obtains $(r_0/\ra)^{7/2}\rightarrow 2$, and the dimensionless torque $n_\omega$ approaches a positive value that depends on $r_{\rm out}/r_0$ and $\chi$.
If accretion is weak and  $r_{\rm out}\rightarrow \infty$, then
\begin{equation}
\label{eq.tau_star_eta1}
\tau_\star   \approx - \frac{\mu^2}{\rc^3} \, \left(\frac{\cos^2{\chi}}{3} + \frac{\alpha \delta_0 \sin^2{\chi}}{4} \right)  .
\end{equation} 

The spin-up and spin-down dimensionless torques for $\chi\rightarrow 0$ are given in Table~\ref{tab.nw}.
{ For $\eta > 0$, $f_\star = 1$ and $\varkappa' = \varkappa$, the inner-edge terms $\varkappa f_P$ cancel in Eq.~\eqref{full_nw_subst}, and for $r_{\rm out}\rightarrow\infty$ the condition $n_\omega = 0$ gives
\begin{equation}\label{eq.omega_eq}
    \omega_{\rm eq} = \frac{6f_\chi + \cos^2\chi}{6f_\chi + 2\cos^2\chi}\, ,
\end{equation}
independent of $\eta$: $\omega_{\rm eq} = 0.875$ ($0.962$), $0.915$ ($0.976$), $0.948$ ($0.986$) and $0.976$ ($0.994$) for $\chi = 0^\circ$, $30^\circ$, $45^\circ$ and $60^\circ$ ($\delta_0 = 0.05$), where the values in parentheses are obtained with $r_{\rm out} = 2^{2/3}\rc$.  
In \citet{GL79b}, the magnetic pitch grows outwards $\propto r^{3}$, resulting in a larger spin-down contribution than in our model and a low equilibrium fastness, $\omega_{\rm eq}\approx0.35$.  This prescription for the magnetic field is argued to be an overestimation~\citep[see, e.g.][]{Bozzo09}.
In our model the toroidal field winds up more slowly: its pitch grows $\propto r^{3/2}$ for $r \gg \rc$, and the spin-down contribution converges.
}

\begin{figure}
  \centering
  \includegraphics[width=\hsize]{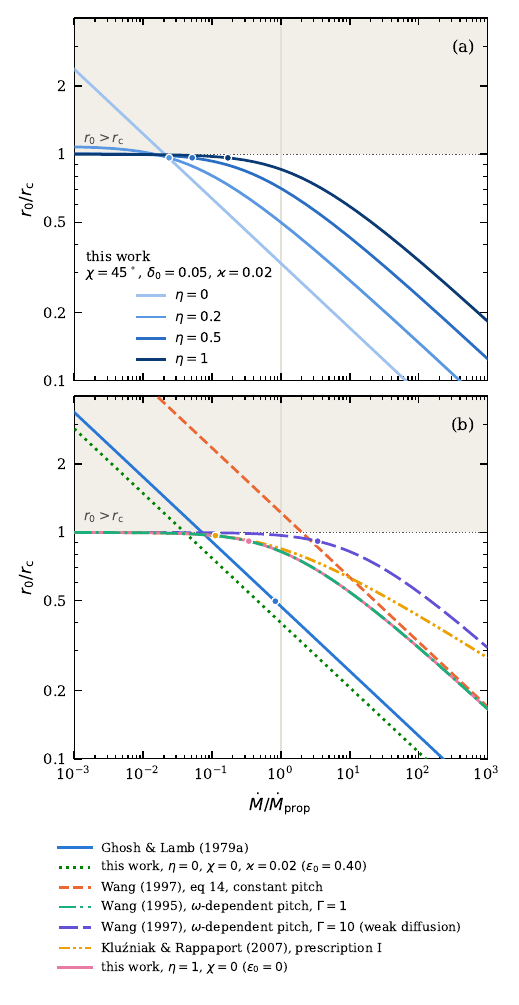}
  \caption{Inner disc radius in units of $\rc$ as a function of $\Mdot/\Mdotpr$, which makes the curves independent of $\mu$, $M$ and $P_{\rm s}$.
(a) Eq.~\eqref{eq.r0_new} for $\eta = 0$, $0.2$, $0.5$ and $1$, with $\chi = 45^\circ$, $\delta_0 = 0.05$ and $\varkappa = 0.02$.
(b) Comparison of different models from the literature, as indicated in the legends. Dots mark zero torque on the star for $r_{\rm out} \rightarrow \infty$.
The Wang models use $\eta = 1$ and $\chi = 0^\circ$.
For threaded models a constant change in the toroidal-field amplitude (change of $\Gamma$) shifts a curve horizontally. The green dash-dot curve overlays the pink solid curve, as well as the equilibrium dot; this shows that our solution for    $\eta = 1$ and $\chi = 0^\circ$ is identical to that of \citet{Wang95}.
}
  \label{fig:r0comp}
\end{figure}

For a diamagnetic disc, that is, for $\eta=0$, and for any values of the fastness $\omega$ and magnetic inclination angle $\chi$,  we obtain the following torque acting on the star:
\begin{equation}
\tau_{\star} \approx  f_\star\dot M \sqrt{G\,M\,r_0} \, -\frac{\mu^2}{r_0^3}  \frac{\varkappa'}{\varkappa} \, \varepsilon_0^{7/2}.  
\end{equation}
{
or  $n_\omega = f_\star - \varkappa'/\varkappa$, and its sign is set by the approximate boundary coefficients rather than by $\omega$.

Figure~\ref{fig:r0comp}, lower panel, compares Eq.~\eqref{eq.r0_new} with other prescriptions for $r_0$. The diamagnetic disc ($\eta = 0$, the lowest dotted line) and the model of \citet{GL79a}, the blue solid line, both give $r_0 \propto \ra \propto \Mdot^{-2/7}$, with $r_0 = 0.40\,\ra$ and $0.48\,\ra$, respectively. The models however differ in the torque: in the model of \citet{GL79b}, the field threads the disc
and $\omega_{\rm eq}\approx0.35$ (marked by the dot), whereas for $\eta = 0$ the torque does not depend on $\omega$. Another caveat is that there is no steady state  in the Ghosh–Lamb model for $\omega \gtrsim 0.95$.

All prescriptions with a toroidal field proportional to the shear at $r_0$ give $r_0 \rightarrow \rc$ as $\Mdot$ decreases.
A constant change in the toroidal-field amplitude shifts a curve along $\Mdot$, so that the absolute $\Mdot$ scale is model-dependent, whereas the position of the zero-torque point relative to $\rc$ is not. 
}

\section{Joule heating in disc} \label{sec.joule}

The Joule heating term, $(\textrm{d}Q /\textrm{d}z)_{\rm J} = \frac{\eta_{\rm T}}{4\pi} (\pd{b}{z})^2$, could be included in the energy transfer equation, namely the last equation in \eqref{vert_1}. Here, $\eta_{\rm T}$ is the turbulent magnetic diffusivity coefficient \citep{CH98a, Rekow2000, OL01}. To compare this term with the mean viscous heating rate, $(\textrm{d}Q /\textrm{d}z)_{\rm vis} \sim Q_0 / z_0$, we recall that the viscous heat flux per unit surface area, $Q_0$, is given by Eq.~\eqref{eq.Q0}. Assuming $\eta_{\rm T} = \nu_{\rm T} \sim \alpha \omegak r^2 \delta^2$ (see Eq.~\eqref{skin}),  $b = b^\mathrm{u} \, z/z_0$, and that $\Wrf$ is given by Eq.~\eqref{eq.wrf_wang}, we obtain the following estimate:
\begin{equation}\label{eq.joule_r}
    \frac{(\textrm{d}Q /\textrm{d}z)_{\rm J}}{(\textrm{d}Q /\textrm{d}z)_{\rm vis}} = \frac{2\alpha}{3} \delta \frac{(1 - (r/\rc)^{3/2})^2 \times \textrm{Threshold}}{(r/r_0)^3 g(r/r_0, \eta) + (r/\ra)^{7/2} f_{1/2}(r/r_0))},
\end{equation}
where Threshold $= 1$ in the region where the toroidal field is induced and Threshold $= 0$ outside this region. 
This relation can become of order unity only very close to the inner disc radius when $\eta\sim1$.

Following the same reasoning as  \citet{KR07}, we do not take the Joule heat into consideration and take into account only the heat produced \akglOK{locally} by viscous processes, since it is not clear which part of Joule heat is released outside the disc. Future models of magnetized discs should account for the Joule heat, possibly treating it as a non-local source of heat.

\end{appendix}
\end{document}